\documentclass[finalnew]{agujournal2018arXiv}
\journalname{JGR-Solid Earth}

\newcommand*\patchAmsMathEnvironmentForLineno[1]{
  \expandafter\let\csname old#1\expandafter\endcsname\csname #1\endcsname
  \expandafter\let\csname oldend#1\expandafter\endcsname\csname end#1\endcsname
  \renewenvironment{#1}
     {\linenomath\csname old#1\endcsname}
     {\csname oldend#1\endcsname\endlinenomath}}
\newcommand*\patchBothAmsMathEnvironmentsForLineno[1]{
  \patchAmsMathEnvironmentForLineno{#1}
  \patchAmsMathEnvironmentForLineno{#1*}}
\AtBeginDocument{
\patchBothAmsMathEnvironmentsForLineno{equation}
\patchBothAmsMathEnvironmentsForLineno{align}
\patchBothAmsMathEnvironmentsForLineno{flalign}
\patchBothAmsMathEnvironmentsForLineno{alignat}
\patchBothAmsMathEnvironmentsForLineno{gather}
\patchBothAmsMathEnvironmentsForLineno{multline}
}

\usepackage{upgreek}
\linenumbers*[1]
\usepackage{graphicx} 
\usepackage{url}
\usepackage{amsmath}
\usepackage{color}
\usepackage{times}
\usepackage{hyperref}
\hypersetup{
    bookmarks=true,         
    unicode=false,          
    pdftoolbar=true,        
    pdfmenubar=true,        
    pdffitwindow=false,     
    pdfstartview={FitH},    
    pdftitle={My title},    
    pdfauthor={Author},     
    pdfsubject={Subject},   
    pdfcreator={Creator},   
    pdfproducer={Producer}, 
    pdfkeywords={keyword1, key2, key3}, 
    pdfnewwindow=true,      
    colorlinks=true,       
    linkcolor=magenta,          
    citecolor=magenta,        
    filecolor=magenta,      
    urlcolor=cyan           
}

\graphicspath{{./jpg/}}
\usepackage[ddmmyyyy,hhmmss]{datetime}

\setkeys{Gin}{draft=false}

\newcommand{\brac}[1]{\left \{ #1 \right \}}

\usepackage[finalnew]{trackchanges}
\addeditor{KO}

\begin{document}


\nolinenumbers


\title{Dynamics, radiation and overall energy budget of earthquake rupture with coseismic off-fault damage}


\authors{Kurama Okubo\affil{1,2*}, Harsha S. Bhat\affil{2}, Esteban Rougier\affil{3}, Samson Marty\affil{2}, Alexandre Schubnel\affil{2}, Zhou Lei\affil{3}, Earl E. Knight\affil{3}, Yann Klinger\affil{1}}

\affiliation{1}{Institut de Physique du Globe de Paris, Sorbonne Paris Cit\'e, Universit\'{e} Paris Diderot, UMR 7154 CNRS, Paris, France.}
\affiliation{2}{Laboratoire de G\'eologie, \'Ecole Normale Sup\'erieure/CNRS UMR 8538, PSL Research University, Paris, France.} 
\affiliation{3}{EES-17 - Earth and Environmental Sciences Division, Los Alamos National Laboratory, New Mexico, USA}
\affiliation{^*}{Currently working at Earth and Planetary Sciences, Harvard University, Cambridge, MA, USA.}

\correspondingauthor{K. Okubo}{kurama\_okubo@fas.harvard.edu}
  
\begin{keypoints}
\item \change{Coseismic off-fault damage is activated by earthquake ruptures, whose feedback plays an important role in rupture dynamics.}{Earthquake ruptures dynamically activate coseismic off-fault damage, whose feedback plays an important role in rupture dynamics.}
\item We show the mechanism of dynamically activated off-fault fractures, and its effect on rupture velocity and enhanced high-frequency radiation. 
\item The contribution of off-fault damage to the overall energy budget associated with earthquakes is non-negligible even at depth.
\end{keypoints}

\abstract{Earthquake ruptures dynamically activate coseismic off-fault damage around fault cores. Systematic field observation efforts have shown the distribution of off-fault damage around main faults, while numerical modeling using elastic-plastic off-fault material models has demonstrated the evolution of coseismic off-fault damage during earthquake ruptures. Laboratory scale micro-earthquake experiments have pointed out the enhanced high-frequency radiation due to the coseismic off-fault damage. However, the detailed off-fault fracturing mechanisms, subsequent radiation and its contribution to the overall energy budget remain to be fully understood because of limitations of current observational techniques and model formulations. Here, we constructed a new physics-based dynamic earthquake rupture modeling framework, based on the combined finite-discrete element method (FDEM), to investigate the fundamental mechanisms of coseismic off-fault damage, and its effect on the rupture dynamics, the radiation and the overall energy budget. We conducted a 2-D systematic case study with depth and showed the mechanisms of dynamic activation of the coseismic off-fault damage. We found the decrease in rupture velocity and the enhanced high-frequency radiation in near-field due to the coseismic off-fault damage. We then evaluated the overall energy budget, which shows a significant contribution of the coseismic off-fault damage to the overall energy budget even at depth, where the damage zone width becomes narrower. The present numerical framework for the dynamic earthquake rupture modeling thus provides the insight into the earthquake rupture dynamics with the coseismic off-fault damage.}

\clearpage
\section{Introduction}

Coseismic off-fault damage has been recognized as a key factor towards understanding the mechanisms of dynamic earthquake ruptures and the associated overall energy budget. \citet{sibson1977} conceptually proposed a formulation for the overall energy budget of dynamic earthquake ruptures; a part of the energy released from accumulated strain energy by interseismic deformation is converted to seismic wave radiation, whereas the rest is expended in inelastic deformation processes within fault zone. \citet{wallace1986} then characterized the structure of fault zones from the observation of deep mines in North America, where fault cores are surrounded by fractured rock. \citet{chester1986, chester1993} also proposed similar fault zone structures based on field observations of San Gabriel and Punchbowl faults in southern California. 

Figure \ref{fig:hierarchicalfaultsystem} illustrates the schematic of a hierarchical fault structure across length scales ranging from regional fault systems to microfractures. The geometrical complexity of fault system is usually discussed in kilometric scale (Figures \ref{fig:hierarchicalfaultsystem}a and \ref{fig:hierarchicalfaultsystem}b). However, when focusing on a part of a fault system, smaller scale fracture networks are observed around faults after the earthquake rupture propagates on the main faults (Figure \ref{fig:hierarchicalfaultsystem}c). These mesoscopic off-fault fractures also have an effect on the displacement field around the faults \citep{manighetti2004,cappa2014}.
Eventually, Figure \ref{fig:hierarchicalfaultsystem}d shows the fault zone structure involving microscopic fractures around the fault core. Field measurements of the microfracture density as a function of distance in fault-normal direction have been conducted in order to understand the spatial distribution and geometric characteristics of the off-fault damage zones \citep{shipton2001,mitchell2009,faulkner2011,savage2011}. \citet{mitchell2012a} showed that the microfracture density is significantly higher close to the fault and exponentially decreases with distance from the fault core, evidencing the presence of coseismic off-fault damage in microscale (Figures \ref{fig:hierarchicalfaultsystem}e and \ref{fig:hierarchicalfaultsystem}f). Since all these geometrical complexities of fractures in a wide range of length scale play a role in the faulting process, the modeling of coseismic off-fault damage is crucial to better understand the rupture dynamics, the radiation and the overall energy budget associated with earthquakes. 

Numerous studies have been performed via theoretical approaches, experiments and numerical modeling to evaluate the effect of coseismic off-fault damage on the earthquake ruptures. \citet{poliakov2002} and \citet{rice2005} showed the potential failure area around rupture front with steady-state cracks \add[KO]{ and pulses} based on theoretical formulations. 
\citet{marty2019} performed laboratory experiments of labo-scale dynamic ruptures with saw-cut rock specimens. They found enhanced high-frequency radiation in acoustic recordings during stick-slip events, considered to be effected by the coseismic off-fault damage, which is of great interest for understanding the high-frequency components in near-field ground motion \citep{hanks1982,castro2013}.

The numerical modeling of coseismic off-fault damage has been also conducted to demonstrate the evolution of the off-fault damage activated by dynamic earthquake ruptures and its effect on the rupture dynamics \citep{yamashita2000, dalguer2003, andrews2005, benzion2005, ando2007b, templeton2008, viesca2008, ma2010, dunham2011a, bhat2012, gabriel2013,  thomas2018a}. 
However, up to now state-of-the-art numerical techniques used for earthquake rupture modeling were not able to describe detailed off-fault fracturing processes as actual tensile and shear (Mode I and Mode II) fractures mainly due to limitations of computation and model formulations. Hence the role of coseismic off-fault damage activated by the dynamic earthquake ruptures remains to be fully understood. Therefore, our aim in this paper is to model the activation of off-fault fracture networks by dynamic earthquake ruptures to evaluate the effect of coseismic off-fault damage on the rupture dynamics, the radiation and the overall energy budget.

We used the combined finite-discrete element method (FDEM) to model the dynamic earthquake rupture with the coseismic off-fault damage. It allows for the activation of both off-fault tensile and shear fractures based on prescribed cohesion and friction laws so that we can quantify the effect of coseismic off-fault damage on the rupture dynamics, the radiation and the overall energy budget. 

We firstly demonstrate the 2-D dynamic earthquake rupture modeling with coseismic off-fault damage. We then show the mechanisms of secondary off-fault fractures, and its effect on the rupture velocity and the radiation. Eventually, we calculate the evolution of energy components associated with the dynamic earthquake rupture to investigate the overall energy budget.

\section{Dynamic earthquake rupture modeling with coseismic off-fault damage}
We performed the dynamic earthquake rupture modeling with a planar strike-slip fault in plane strain condition, surrounded by the intact rock, allowing for the activation of off-fault fractures. Figure \ref{fig:modeldescription}a shows the model description for the 2-D dynamic earthquake rupture modeling. The rupture is artificially nucleated from the nucleation patch, where the peak friction is lower than the initial shear traction on the main fault. The size of nucleation patch $L_c$ is determined by the critical crack length \citep[][]{palmer1973}.
Then it propagates bilaterally on the main fault, dynamically activating off-fault fractures. The x axis is along the fault-parallel direction, while the y axis is along the fault-normal direction. Figure \ref{fig:modeldescription}b shows the schematic of case study with depth. We performed a set of 2-D dynamic earthquake rupture modeling to investigate the evolution of coseismic off-fault damage and its effect with depth. The z axis is thus along depth. We conducted 2-D simulations at every 1km from z = 2km to 10km depth with corresponding initial stress state as shown in Figure \ref{fig:modeldescription}c. We assume lithostatic condition with depth so that the confining pressure linearly increases with depth. The quasi-static process zone size $R_0$ (see eq. \ref{eq:processzonesizeconstGIIC}) decreases with depth when the fracture energy on the main fault $G_{IIC}^f$ is kept constant with depth (Figure \ref{fig:modeldescription}c). \add[KO]{Note that the case study does not address the 3-D effect (e.g. free surface) as we model the dynamic ruptures in plane strain condition.}

For the sake of fair comparison between different depths, the model parameters are nondimensionalized by scaling factors. $R_0$ [m] and shear wave velocity $c_s$ [m/s] are used to scale the length [m] and the time [s] by $R_0$ and $R_0/c_s$, respectively. Subsequently, other variables are also nondimensionalized by the combination of those two scaling factors. Since the density of medium does not change during simulations, the nondimensionalization of mass is not necessary in our problem.

The methodology of  the combined finite-discrete element method (FDEM) is described in Appendix \ref{sec:appa}. More details of the numerical framework to model dynamic earthquake rupture with FDEM can be found in \citet{okubo2018h}. 
The parameters used for the case study with depth are summarized in Table \ref{tab:parameters}.
We used FDEM-based software tool, Hybrid Optimization Software Suite - Educational Version (HOSSedu) for the dynamic earthquake rupture modeling \citep{hoss2015}. Before modeling dynamic earthquake ruptures with coseismic off-fault damage, we conducted the cross-validation of the FDEM using purely elastic medium to assess the achievable accuracy of earthquake rupture modeling. The results are summarized in Appendix \ref{sec:appb}.

Figure \ref{fig:snapshotSR} shows a snapshot of dynamic earthquake rupture with dynamically activated off-fault fractures, where particle velocity field and the fracture traces around the main fault are superimposed. The seismic ratio S is equal to 1.0 (see eq. \ref{eq:Sratio}), which results in the sub-Rayleigh rupture during the simulation with the coseismic off-fault damage. The off-fault fractures are plotted when the traction applied on the potential failure plane (i.e. boundary of meshes) reaches the cohesive strength and the cohesion starts weakening. Bottom and left axes indicate the fault-parallel and fault-normal distance in physical length scale, while top and right axes indicate the nondimensionalized length scale.

The off-fault fractures are initiated around the rupture tip, and then it forms an intricate fracture network as the main rupture propagates on the main fault. The particle velocity field is significantly perturbed due to the coseismic off-fault damage. The extensional side of the main fault is mostly damaged, which is supported by the theoretical analysis of potential failure area \citep{poliakov2002,rice2005} and other simulations \citep[e.g.][]{andrews2005}. The off-fault fractures form an intricate network by means of fracture coalescence, comprised of tensile, shear and mixed mode fractures. We later discuss this off-fault fracturing process \add[KO]{under a relatively steep angle of the maximum compressive principal stress $\sigma_1$ to the fault ($\psi$ = 60$^{\circ}$}), and its effect on the radiation in near-field and the overall energy budget. 

Figure \ref{fig:snapshotSS} shows a set of snapshots for the supershear case with S = 0.7. The rupture is nucleated and propagates with sub-Rayleigh in the earlier phase. Then a daughter crack is born ahead of the rupture front at T = 4.7 s, which then transitions to supershear rupture. During the rupture transition from sub-Rayleigh to supershear, characteristic damage pattern appears; there is a gap of coseismic off-fault damage around the transition phase (around x = 12km in Figure \ref{fig:snapshotSS}). This characteristic damage gap has been also pointed out by \citet{templeton2008} and \citet{thomas2018a}. \add[KO]{This can be explained by the Lorentz contraction of the dynamic process zone size $R_f(v_r)$ (see \protect\ref{sec:processzonesize}). The dynamic process zone size asymptotically shrinks at the limiting speed of the rupture. Since the damage zone size is approximated as a same order of the process zone size, the damage gap appears during the supershear transition. Then it resumes the off-fault fracturing after the initiation of supershear rupture.} \add[KO]{We need to further explore the transition of stress concentration associated with the dynamic rupture during supershear transition in order to better explain the mechanism of damage gap.}

\section{Mechanism of coseismic off-fault damage}
\label{sec:fracmechanics}
We first investigate the fracturing process in off-fault medium activated by the dynamic rupture propagation. We aim to show how the off-fault fracture network evolves as the dynamic earthquake rupture propagates on the main fault. Figure \ref{fig:fracmechanism} shows the traces of tensile (dilating), shear and mixed mode fractures in the off-fault medium at two time steps replotted from Figure \ref{fig:snapshotSR}. 
To highlight the potential failure area, the first stress invariant normalized by its initial value $I_1(t)/I_1^{\text{init}}$ (eq. \ref{eq:I1}) for tensile fractures and the normalized closeness to failure  $d_{MC}/d_{MC}^\text{init}$ (eq. \ref{eq:dMC}) for shear fractures are respectively superimposed on the traces of secondary fractures (see \ref{app:potentialfailurearea}). Note that \remove[KO]{the} both regions do not assure the traction reaches to the peak cohesion due to the threshold for plotting. Thus the potential failure planes in the regions are not necessarily broken.

The intricate fracture network is formed even after the dynamic earthquake rupture passes because the stress concentration still remains behind the rupture front due to the internal feedback from the off-fault fracture network itself. The stress concentration is then relaxed by the activation of new fractures in the off-fault medium.
The tensile fractures are always initiated just behind the rupture front with a certain dominant orientation. This dominant orientation is experimentally and theoretically studied by \citet{ngo2012}, and has a reasonable correspondence with the orientation obtained from our analysis. It is also remarkable that the position of fracture initiation is scaled by the dynamic process zone size $R_f(v_r)$ (see \ref{sec:processzonesize}), which is also pointed out by \citet{viesca2009}.

We then examined a set of case study with depth to investigate the evolution of damage pattern, fracture density and the damage zone width with depth. Figure \ref{fig:fracmechwithdepth} shows the traces of off-fault fracture network and the spatial distribution of fracture density at 2km, 6km and 10km depths. The isolated fracture network, in which all fractures coalesce with each other, is separately plotted with different colors. 
The dimensions are scaled by $R_0$ so that the size of the fracture network \change[KO]{are}{is} visually comparable. The number of isolated fracture network is more for the shallower case than the deeper case, implying the off-fault fracture network becomes more intricate and denser with depth. 
To evaluate the distribution of fracture density, we firstly imposed representative square grids around the fault as shown in Figure \ref{fig:fracmechwithdepth} at 2km depth, and calculated the normalized fracture density $\hat{P}_{21}$ in the each grid defined as
\begin{equation}
\hat{P}_{21} = \dfrac{\text{Length of fracture trace in a grid}}{\text{Area of grid}} R_0.
\end{equation}
We carefully chose the grid size, which involves a reasonable number of potential failure planes. In this analysis, the grid size is set as 0.2$R_0$, which is, on average, three times larger than the size of potential failure planes. For the sake of comparison between different depths, the magnitude of $\hat{P}_{21}$ is normalized by its maximum value at 10km depth. We \remove[KO]{then} found that the fracture density globally increases with depth, though it does not monotonically increase due to complicated internal feedback in the off-fault fracture network. \add[KO] {The increase in the normalized fracture density with depth enhances the contribution of coseismic off-fault damage to the overall energy budget, discussed in section} \ref{sec:overallenergybudget}.
Note that the pulverization in the vicinity of main fault is not modeled because of limitations in the size of the potential failure planes. It would be resolved by incorporating with constitutive damage models \citep{bhat2012}.

Figure \ref{fig:rosediagram} shows the rose diagram showing the orientation of off-fault fractures. The size of bars  \add{in the rose diagram} is normalized by the sum of all types of fracture. It has a dominant orientation for tensile fractures, which corresponds to the orientation of $\sigma_1$. The shear fractures also have two dominant orientations, which correspond to the conjugate failure planes inferred from the Mohr-Coulomb failure criterion. There is no dominant orientation for mixed mode fractures. The fraction of the each type of fracture shows that the population is fairly balanced, whereas the fraction of tensile fracture decreases with depth because more intricate fracture network is formed at depth.

Figure \ref{fig:damagezonewidth} shows the evolution of the damage zone width with depth. The damage zone is inferred from the envelope of secondary fracture network \add[KO]{at the scaled rupture length} $x = 5R_0$, $10R_0$ and $20R_0$ \add[KO]{to compare at the same stage of dynamic ruptures with depth, and to investigate representative damage zone width associated with the in-situ confining pressure}. Since there are few off-fault fractures being activated \change{in}{on} the compressional side, we only plot the damage zone width on \add[KO]{the} extensional side. The damage zone width follows, up to a constant factor, the quasi-static process zone size. Hence the damage zone width decreases with depth, forming the flower-like structure, with fracture connectivity increasing with depth.
This structure of damage zone with depth is in agreement with the observations \citep[e.g.][]{cochran2009}.

We here demonstrated the fracturing process of off-fault medium activated by the dynamic earthquake rupture with depth. The dynamic activation of off-fault fracture network has an effect on the rupture dynamics and causes additional radiation, which effects high-frequency components in near-field ground motion discussed in the following section. We also examined the mesh dependency \change[KO]{for}{of} the fracturing process because the potential failure planes are restricted to the element boundary, which is discussed in Appendix \ref{sec:appc}.

\section{Rupture velocity}
We next focus on the rupture velocity on the main fault. Figure \ref{fig:xtplot} shows the evolution of slip velocity on the main fault with four cases; S = 1.0 or 0.7 at 2km depth, each of which with or without off-fault damage. For the cases without off-fault damage, the activation of secondary fracture is suppressed by the extremely high cohesion for both tensile and shear fractures. Here, we plot the contour of slip velocity in space and time. In Figure \ref{fig:xtplot}a, there is a clear transition from sub-Rayleigh to supershear around $x/R_0 = 20$, which is also shown in the inset. 
However, when the coseismic off-fault damage is taken into account, the supershear transition is not observed during the simulation as shown in Figure \ref{fig:xtplot}b. Hence, the secondary fractures can arrest, or delay, supershear transition in a certain stress conditions. This can be explained by the increase in critical slip distance due to the coseismic off-fault damage. The supershear transition length $L^{\text{trans}}$ can be estimated from the Andrews' result \citep{andrews1985,xia2004} as following
\begin{equation}
L^{\text{trans}} = \dfrac{1}{9.8(S_{\text{crit}} - S)^3} \dfrac{1+\nu}{\pi}\dfrac{\tau^p - \tau^r}{(\tau - \tau^r)^2} \mu D_c,
\label{eq:sstransition}
\end{equation}
where $S_{\text{crit}}$ is the threshold for the supershear transition ($S_{\text{crit}} = 1.77$ for 2-D), $\nu$ is Poisson's ratio, $\tau_p$, $\tau_r$ and $\tau$ are peak friction (eq. \ref{eq:peakfric}), residual friction (eq. \ref{eq:residualfric}) and shear traction on the fault, respectively, $\mu$ is shear modulus and $D_c$ is critical slip distance for friction (eq. \ref{eq:Dcwithdepth}). $D_c$ is initially uniform on the main fault. However, the effective critical slip distance $D_c^{\text{eff}}$ \remove{(eq. \protect\ref{eq:critDc})}, which takes into account the energy dissipation in the off-fault medium due to the coseismic off-fault damage, increases with the rupture length as discussed in later section \change{(Figure \protect \ref{fig:effectiveDc})}{(Section \protect\ref{sec: Dceff})}. Therefore, $L^{\text{trans}}$ also increases due to the coseismic off-fault damage as it is proportional to $D_c$. 

Figures \ref{fig:xtplot}c and \ref{fig:xtplot}d show the cases with $S=0.7$, where the rupture transitions to supershear for both cases with and without off-fault damage because of the large contrast of the initial shear traction to the normal traction on the main fault. The time of supershear transition is delayed with off-fault damage due to the decrease of rupture velocity, whereas the difference of transition length is still obscure with these results. The two insets in the figures show the clear difference in the peak of slip velocity and the fluctuation. In addition, the rupture arrival is delayed by the coseismic off-fault damage, implying the decrease in rupture velocity.

The rupture velocity is calculated from first arrival times along the main fault. Figure \ref{fig:rupturevelocity} shows the evolution of rupture velocity in time. We take the time derivatives of first arrival time in discretized space along the main fault to calculate the representative rupture velocity at a certain position. Since it is difficult to capture the exact time when rupture velocity jumps to supershear, where the curve of first arrival time has a kink and is non-differentiable, the error caused by the smoothing of the rupture velocity is taken into account as shown by the error bars in Figure \ref{fig:rupturevelocity}. Therefore, the markers in the forbidden zone $c_R < v_R <c_s$ do not conclusively indicate that the rupture velocity is between them due to the uncertainty. 

Regardless of the uncertainty, the comparison between the cases with and without off-fault damage shows the effect of coseismic off-fault damage on the rupture velocity and the supershear transition. The rupture transitions to supershear for both cases with S=0.7, \change[KO]{though}{whereas} the rate of increase in rupture velocity is lower for the case with \add[KO]{the off-fault} damage. However, the supershear transition is suppressed due to the coseismic off-fault damage with S=1.0. Further parametric study would narrow down the criteria of supershear transition and \change[KO]{evaluate the change in}{would provede} supershear transition length.

\section{High-frequency radiation in near-field}
The origin of high-frequency radiation has been studied over decades \citep[e.g.][]{madariaga1977a,hanks1981, hanks1982, ohnaka1987, dunham2011b, castro2013, passelegue2016b,marty2019}. There are multiple factors that effect the high-frequency radiation, such as sudden nucleation and arrest of rupture, complex fault geometry, roughness of the fault surface and the nonlinear response in subsurface sedimentary rock. In this study, we propose that the coseismic off-fault damage is also a candidate which effects the high-frequency radiation. 
\change[KO]{Since the intricate fracture network is dynamically formed during the rupture propagation, the radiated wave field in near-field is significantly perturbed by the additional radiation from the secondary fractures as well as the oscillation of the slip velocity on the main fault.}{When the secondary fracture is activated in the off-fault medium, it behaves as a secondary source to radiate waves, which contributes to the enhancement of high-frequency components in the near-field ground motion. The off-fault fracture network also causes the scattering wave due to the structural heterogeneities.} 

Figure \ref{fig:waveform}a shows the waveforms of the fault-normal acceleration at $x = 12.4R_0$, 2km depth with S=1.0. The amplitude is compressed to highlight the signals arising from the coseismic off-fault damage. The theoretical P and S wave arrival time and the rupture arrival time at $x = 12.4R_0$ are indicated in Figure \ref{fig:waveform}a. After the P and S wave arrival, there is a well-aligned signal around 6s, which is caused by the stress perturbation around the rupture front. 
\change[KO]{Then there is significant spikes in near-field ground acceleration, which is the signal from the secondary fracturing.}{Significant high-frequency spikes then arise after the main rupture arrival, which are caused by the secondary off-fault fracturing, instead of the regular near-field radiation from the main rupture.} These spikes are observed up to 2$R_0$ from the fault, corresponding to the damage zone width \add[KO]{at this rupture length}. Figure \ref{fig:waveform}b shows the spectrogram of the near-field ground acceleration ($y=-0.5R_0$). The spikes are observed at t = 8.7s and 9.9s even after the passage of rupture on the main fault due to the secondary fracturing activated by the internal feedback of off-fault fracture network.

We then investigate the spatial distribution of the high-frequency radiation with depth using the critical frequency $f^{\text{crit}}$, where the amplitude spectrum decays from the mean level of low-frequency band. Figure \ref{fig:freqwithdepth} shows the spatial distribution of $f^{\text{crit}}$, and the near- and far-fault spectra. Note that the far-fault does not mean the far-field ground motion, where the near-field and intermediate-field terms are negligible in point source model. \add[KO]{The near-fault amplitude spectrum in the right column of Figure \protect \ref{fig:freqwithdepth} is evaluated within the damage zone, which is not intended for the direct comparison to the observations as there might be technical issues on the implementation of instruments close to the fault.}

\remove[KO]{Note that the results in this section are in the context of near-field ground motion. We recorded fault-normal velocity over the plotting area.} The signal time window starts from the first arrival time at the location to the end of simulation. We applied a band-pass filter of 0.1-100 Hz and a Tukey window\remove[KO]{ in time domain}. The spectra for the case without off-fault damage are superimposed at 2km depth for the comparison. The results show significant high-frequency radiation \add[KO]{caused by the secondary fractures}, \change[KO]{whose extent is more than the damage zone width.}{which propagates even outward from the damage zone.} \add[KO]{Therefore, although} the high-frequency radiation is quickly attenuated due to the geometric dispersion, the coseismic off-fault damage is clearly one of the factors which effects the high-frequency radiation in \change[KO]{near-field}{the near field}. 

The enhanced high-frequency radiation associated with dynamic ruptures is also studied by experiments. \citet{marty2019} conducted systematic stick-slip experiments with saw-cut Westerly granites under servo-controlled triaxial loading with the confining pressure $\sigma_3$ ranging from 10 MPa to 90 MPa to investigate the enhanced high-frequency radiation in acoustic recordings of the stick-slip events.
The acoustic sensors are externally located on the surface of specimen, which record the motion of the normal component to the surface. The representative Fourier spectra are obtained by taking an average of 13 acoustic sensors on the specimen to get rid of directivity effect. Further information of these experiments can be found in \citet{marty2019}.

Figure \ref{fig:experiments}a shows the Fourier spectra with different confining pressures. \change[KO]{For the sake of comparison, each spectral amplitude is normalized by each value at low frequency.}{Each spectrum amplitude is normalized by its corresponding stress drop in order to compare the high-frequency content.} The theoretical critical frequencies for $v_r$=2000m/s (for sub-Rayleigh) and $v_r$=5000m/s (for supershear) are indicated, implying the rupture transitions to supershear with $\sigma_3 \geq 20$MPa. Certainly, one of the possible reasons for the enhanced high-frequency components is the supershear transition. However, there is also an enhanced frequency band from 400 kHz to 800 kHz, which can be caused by the coseismic off-fault damage. Thus they conducted back-projection analysis to investigate the spatiotemporal evolution of seismic energy release in this frequency band.

Figure \ref{fig:experiments}b \remove{the} shows snapshots of back-projection results for a certain stick-slip event with $\sigma_3 = 90$ MPa. The color contour \change[KO]{shows}{indicates} the normalized coherency function, which \change[KO]{indicates}{shows} the most likely location of the origin of the signal within \change[KO]{this}{the} frequency band.  The rupture is spontaneously nucleated at the edge of the saw-cut surface, and propagates downward. The theoretical rupture front is also superimposed on the fault surface. The results show that the high-frequency signals within this band originate just behind the rupture front, which can be caused by the coseismic off-fault damage as compared to the off-fault fracturing process discussed in section \ref{sec:fracmechanics}. The experimental results demonstrate the first-order analysis of the mechanism of enhanced high-frequency radiation, which is in agreement with the secondary fracturing mechanism  \change{inferred by}{as shown in } \change{the dynamic earthquake rupture modeling with coseismic off-fault damage.}{Figure \protect \ref{fig:fracmechanism}}

\section{Overall energy budget}
\label{sec:overallenergybudget}
The overall energy budget of an earthquake event plays a key role in understanding the characteristics of the earthquake source, the change of potential energy and the radiation. Here, we first describe the formulation of energy balance, which can be used to \change{estimate}{evaluate} the overall energy budget associated with the dynamic earthquake rupture with coseismic off-fault damage. Although there are various approaches to derive the energy conservation law of earthquake ruptures \citep[e.g.][]{rivera2005,fukuyama2005,shi2008,xu2012a}, we reidentify the energy components in a suitable form for the analysis of the overall energy budget with our numerical framework.  

The overall energy budget is evaluated in an inner volume $V_0$ as shown in Figure \ref{fig:energyschematic}a, which encompassing entire rupture zone and off-fault fractures. Then the energy components associated with the overall energy budget are written as follows:

\begin{itemize}
\item Elastic strain energy
\begin{equation}
\Delta W = \int_{V_0} \left[ \int_0^{\varepsilon_{ij}}  \sigma_{ij}d\varepsilon_{ij}  \right] dV,
\end{equation}
where $\varepsilon_{ij}$ is strain tensor and $\sigma_{ij}$ is stress tensor. Note that the initial strain is defined to be zero, whereas the initial stress is nonzero. This configuration is commonly used in seismology, discussed in \citet[][BOX 8.5]{aki2002}.

\vspace{15pt}
\item Kinetic energy
\begin{equation}
E_K = \int_{V_0} \dfrac{1}{2} \rho \dot{u}_i \dot{u}_i dV,
\end{equation}
where $\rho$ is density and $\dot{u}_i$ is particle velocity.

\vspace{15pt}
\item Radiated energy
\begin{equation}
E_{R}(t)= - \int_0^t dt \int_{S_0}  \left( T_i - T_i^0 \right) \dot{u}_i dS,
\label{eq:ER}
\end{equation}
where $S_0$ is the closed surface of $V_0$,  $T_i$ is the traction on $S_0$ and $T_i^0=T_i(0)$. $E_R$ is essentially the work done by $V_0$ to outer volume $V_1$.
Note that the canonical $E_R$ is determined at the end of earthquake event, where $E_K=0$ in $V_0$ \citep{kostrov1974a}, whereas $E_R$ defined by equation (\ref{eq:ER}) is a function of time due to our model description with infinite fault length, where the rupture does not cease during simulation. 

\vspace{15pt}
\item Fracture energy on the main fault
\begin{equation}
E_{G}^{on} = \int_{\Gamma^{\text{Main Fault}}} \left[ \int_{\delta_{II}^{f,e}}^{\min \brac{D_c^{\text{main}}, \delta_{II}^*}} T_t (\delta_{II}) - \tau_r d\delta \right] dS,
\end{equation}
where $\Gamma^{\text{Main Fault}}$ is the surface of main fault, $\delta_{II}^{f,e}$ is the critical slip for elastic loading of friction (see \ref{sec:failurecriteria}), $D_c^{\text{main}}$ is critical slip distance of slip-weakening law on the main fault, $\delta_{II}^*$ is slip at $t$ and $T_t$ is shear traction on the fault. \add[KO]{Note that $E_{G}^{on}$ is always positive during the slip-weakening of friction.}

\vspace{15pt}
\item Fracture energy associated with the off-fault damage
\begin{equation}
E_{G}^{off} = \sum_i^N \int_{ \Gamma_i^{\text{Off-fault}}} \left[\int_{\delta_{I/II}^{c,e}}^{\min \brac{\delta_{I/II}^{c,c}, \delta_{I/II}^*}} {C_{I/II}(\delta_{I/II}) }d\delta + \int_{\delta_{II}^{f,e}}^{\min \brac{D_c^{\text{off}}, \delta_{II}^*}} T_t (\delta_{II}) - \tau_r d\delta \right] dS,
\end{equation}
where $\Gamma_i^{\text{Off-fault}}$ is the surface of off-fault fracture, $N$ is the number of off-fault fractures, $\delta_{I/II}^{c,e}$ is the critical slip for elastic loading of tensile and shear cohesion, $\delta_{I/II}^{c,c}$ is the maximum displacement for softening of tensile and shear cohesion, $D_c^{\text{off}}$ is critical slip distance of slip-weakening law in the off-fault medium, $\delta_{I/II}^*$ is opening and shear displacement at $t$ and $C_{I/II}$ is the tensile and shear cohesion.

\vspace{15pt}
\item Heat energy
\begin{equation}
 E_H = \int_{\Gamma^{\text{Main Fault}} +  \Gamma_i^{\text{Off-fault}}} \left[ \int_{\delta_{II}^{f,e}}^{\delta_{II}^*} \tau_r d\delta \right] dS.
 \end{equation}

\end{itemize}
\vspace{15pt}

Using the energy components described above, the overall energy budget is written as
\begin{equation}
E_{R} +  E_K + E_{G}^{\text{on}} + E_{G}^{\text{off}} + E_{H} = -(\Delta W + E_{S_0}^0),
\label{eq:ebmaster}
\end{equation}
where
\begin{equation}
E_{S_0}^0 = - \int_0^t dt \int_{S_0}  T_i^0 \dot{u}_i dS.
\end{equation}
$E_{S_0}^0$ originates from the definition of radiated energy in equation (\ref{eq:ER}), which does
not appear in the conventional energy conservation law on earthquake \citep[e.g.][]{rivera2005} because of reasonable approximation processes to estimate the radiated energy. In this study, however, we define the overall energy budget with $E_{S_0}^0$ to rigorously estimate the contribution of each energy components to the overall energy budget. Note that we ignore the consumed energies by elastic loading as they are negligible with large stiffness in cohesion and friction law. The detailed derivation of the overall energy budget can be found in \citet{okubo2018h}.

\subsection{Energy dissipation in off-fault medium}
\label{sec: Dceff}
Figure\remove{s} \ref{fig:energyschematic}b shows the fraction of each energy components in the left side \change{off}{of} equation (\ref{eq:ebmaster}) as a function of rupture length with depth. Each fraction is calculated against $-(\Delta W + E_{S_0}^0)$. Currently, $\Delta W$, $E_{S_0}^0$, $E_{R}$, $E_K$ and $E_{G}^{\text{on}}$ are directly calculated from the simulation, \change{while}{whereas} $E_{G}^{\text{off}}$ and $E_{H}$ are indirectly evaluated with the assumption of average displacement on the off-fault fractures \add{due to the current limitation of post-processing}. \add[KO]{Note that the fracture energy $G_{IIC}^f$ is constant along the main fault.} More details for the calculation of energy components can be found in \citet[][section 3.4]{okubo2018h}.

The fraction of fracture energy associated with the off-fault damage $E_{G}^{\text{off}}$ increases with the rupture length. Moreover, it also increases with depth even though damage zone width becomes narrower at depth. 
To highlight the fraction of $E_{G}^{\text{off}}$ against $E_{G}^{\text{on}}$, we calculated the effective $D_c$ defined as
\begin{equation}
D_c^{\text{eff}} = D_c^{\text{main}} + \dfrac{2E_{G}^{\text{off}}/L}{\tau_p-\tau_r},
\label{eq:critDc}
\end{equation}
where $L$ is the rupture length. Note that we assume unit thickness of the fault. 
Figure \ref{fig:effectiveDc} shows $D_c^{\text{eff}}$ as a function of rupture length with depth calculated from Figure \ref{fig:energyschematic}b. $D_c^{\text{eff}}$ increases with the rupture length and with depth, implying more energy dissipation in the off-fault medium with large ruptures at depth. Up to the half amount of the fracture energy on the main fault can be also dissipated in the off-fault medium due to the coseismic off-fault damage at depth.

\subsection{Seismic efficiency}
Seismic efficiency $\eta_r$ is an important parameter to quantify the proportion of radiated energy to the sum of radiated energy and fracture energy, which essentially evaluates the balance between the radiated energy as seismic waves and the dissipated energy due to on- and off-fault fracturing \citep{kanamori2004}. Since we can only evaluate the temporal radiated energy in equation (\ref{eq:ER}) because of the infinite fault length in our model description, we define modified seismic efficiency for this study, given by
\begin{equation}
\eta_r = \dfrac{E_R + E_K}{E_R + E_K + E_G^{\text{on}} + E_G^{\text{off}}}.
\end{equation}
The physical interpretation of this quantity is same with the canonical seismic efficiency. We evaluated the evolution of $\eta_r$ as a function of rupture length at 2km and 10km depths and compared between the cases with and without off-fault damage to investigate the effect of off-fault damage on the seismic efficiency.
Figure \ref{fig:seismicefficiency} shows the $\eta_r$ for the cases with $S=1.0$ and $S=0.7$. The relative difference between the cases with and without off-fault damage is plotted in insets, defined as
\begin{equation}
\Delta \eta_r = 1 - \dfrac{\eta_r^D}{\eta_r^C},
\end{equation}
where $\eta_r^D$ and $\eta_r^C$ indicate the $\eta_r$ with and without off-fault damage, respectively. There is a significant decrease in $\eta_r$ due to the coseismic off-fault damage, particularly in the deeper case with sub-Rayleigh rupture. This can be explained by the denser and more intricate off-fault fracture network formed with deeper cases. Therefore, although the secondary off-fault fractures effect the additional high-frequency radiation, we maintain that the coseismic off-fault damage absorbs some of available energy, which is inherently converted to the radiated energy for the cases without off-fault damage.

\section{Conclusion}

Our systematic case study with depth demonstrated the mechanisms of coseismic off-fault fracturing and its effect on the rupture dynamics, \add{the} radiation and \add{the} overall energy budget. The damage zone width decreases with depth, whereas the fracture density and the contribution of energy dissipation in off-fault medium globally increases with depth in nondimensional comparison. Overall, Figures \ref{fig:conclusivefigure}a and \ref{fig:conclusivefigure}b show the schematic of fault structure with depth based on this study and the summary of numerical results, inferred from the sub-Rayleigh cases ($S=1.0$).
The fracture density is evaluated using a representative value averaged over space with Figure \ref{fig:fracmechwithdepth}. The damage zone width becomes narrower with depth, whereas the contribution of fracture energy to the overall energy budget rather increases with depth due to the increase in fracture density and complexity of off-fault fracture network.

In this study, we conducted simulations with intact rock, and with fixed orientation of principal stress at 60$^{\circ}$. Therefore, we observed the coseismic off-fault damage only in the extensional side of the fault. However, in nature, the off-fault damage is often observed \change[KO]{in both sides of the fault}{on both sides of the fault}. \add[KO]{Due to the model constraints above, the present results, such as the damage zone width, do not always provide a close quantitative agreement to the observations.} 

The \change{pre-damage}{preexisting damage} of the off-fault medium, \add{the} initial cohesion on the main fault, which is assumed to be zero in this study, and the orientation of the maximum principal stress \change[KO]{might}{also} play a role in the off-fault damage \change{in compressive side}{on the compressional side}.{These need to be investigated by extensive parametric studies.}

The present numerical framework is capable of the application to the natural fault system. \citet{klinger2018} showed the dynamic earthquake rupture modeling on the 2016 Kaik\={o}ura earthquake with the same numerical framework we proposed in this study. They used the dynamic earthquake rupture modeling to resolve the most likely rupture scenario by comparing cosesmic off-fault damage pattern to the observations. 

This study has opened an avenue to model dynamic earthquake ruptures with FDEM, which allows for modeling dynamic earthquake ruptures with coseismic off-fault damage to better understand the fracturing mechanisms, the radiation and the overall energy budget associated with earthquakes.

\appendix
\clearpage
\section{Methodology for modeling coseismic off-fault damage with FDEM}
\label{sec:appa}
Geological faults can be defined as discontinuities in a continuum medium. From this perspective, we consider both the faults and the off-fault damage as an aggregation of fractures at different length scales. FDEM is capable of modeling both continuum deformation and fracturing (i.e. dynamic rupture on the main fault and the off-fault damage) within the same numerical framework. In this appendix, we describe the essence of numerical framework for dynamic earthquake rupture modeling with coseismic off-fault damage. A set of detailed model formulation can be found in \citet{okubo2018h}. More details of main algorithmic solutions used within HOSSedu can be found in a series of monographs \citep{munjiza2004, munjiza2011, munjiza2015}.

\subsection{Initial stress state at depth}
\label{sec:modeldescription}
We follow a similar process to that proposed by \citet{templeton2008} and \citet{xu2012a} to make an assumption of initial stress state as a function of depth. The initial stress state is set for triggering right-lateral strike-slip on the main fault. The initial stress state is uniform in the homogeneous and isotropic elastic medium, given by
\begin{equation} \sigma_{ij}^0 = 
\left[
    \begin{array}{cc}
      \sigma_{xx}^0 & \sigma_{yx}^0 \\
      \sigma_{yx}^0 & \sigma_{yy}^0 
    \end{array}
  \right].
\end{equation}
Note that we assume plane strain conditions. Let normal stress $\sigma_{yy}^0$ on the main fault be given by linear overburden effective stress gradient \add[KO]{as it provides an approximation of the magnitude of normal stress on the main fault with depth}, such that 
\begin{equation}
\sigma _{yy}^0 = -(\rho - \rho_{w}) gz,
\label{eq:syy}
\end{equation}
where $\rho$ is the density of rock, $\rho_{w}$ is the density of water, $g$ is the gravitational acceleration and $z$ is the depth measured from the ground surface. 
The initial shear stress $\sigma_{yx}^0$ is estimated in terms of the seismic S ratio, defined by \citet{andrews1976}, on the main fault such as
\begin{equation}
S = \frac{f_s(-\sigma_{yy}^0) - \sigma_{yx}^0}{\sigma_{yx}^0 - f_d(-\sigma_{yy}^0)},
\label{eq:Sratio}
\end{equation}
where $f_s$ and $f_d$ are the static and dynamic friction coefficients, respectively. The S ratio defines whether the rupture transitions to supershear ($S < 1.77$), or remains sub-Rayleigh ($S > 1.77$) with 2-D purely elastic model (i.e. no off-fault damage).
From equation (\ref{eq:Sratio}), the initial shear stress on the main fault can be written as
\begin{equation}
\sigma_{yx}^0 = \frac{f_s + S f_d}{1+S} (-\sigma_{yy}^0).
\end{equation}
The horizontal compressive stress $\sigma_{xx}^0$ is then determined by the normal stress $\sigma_{yy}^0$, shear stress $\sigma_{yx}^0$  and the given orientation of initial compressive principal stress to the main fault $\psi$ as follows:
\begin{equation}
\sigma_{xx}^0 = \left( 1 - \frac{2\sigma_{yx}^0}{\tan{(2\psi)} \sigma_{yy}^0} \right ) \sigma_{yy}^0.
\end{equation}

\add[KO]{In the present study, we examined the cases with $S=1.0$ and $0.7$ to simulate sub-Rayleigh and supershear ruptures with coseismic off-fault damage, respectively. However, the examined S ratios are relatively low as both $S=1.0$ and $0.7$ lead to supershear transition without the off-fault damage as shown in Figure \protect \ref{fig:rupturevelocity}.} \add[KO]{The analysis of the effect of coseismic off-fault damage with larger S ($>1.5$) remains to be done as the dominant wing cracks are activated from the edges of nucleation patch during nucleation phase, which prevents the rupture nucleation and the propagation along the main fault. Therefore, to avoid the huge stress concentration at the edge of nucleation patch, we need to improve the nucleation process, e.g. a gaussian distribution of peak strength within the nucleation patch.}

\subsection{Damage type and fracture energy}
\label{sec:failurecriteria}
In the FDEM framework, fractures are represented as the loss of cohesion at the interfaces of the finite elements. 
The cohesion and the friction against the opening or shear motion between contactor and target are a function of displacements defined by the aperture $\delta_I$ and the slip $\delta_{II}$ (Figure \ref{fig:app1}a). 
Figure \ref{fig:app1}b shows the mesh discretization and the schematic of off-fault fractures. The cohesive and frictional resistances are applied on every interface between elements, which is regarded as a potential failure plane. Fractures are activated when the cohesion starts to be broken due to the stress concentration of the dynamic earthquake rupture.
Both cohesion and friction curves are divided into two parts, an elastic loading part and a displacement-weakening part as shown in Figure \ref{fig:app1}c and \ref{fig:app1}d. In the elastic loading part, the resistant forces against displacements acting on the interface increase quadratically (for the case of cohesion) or linearly (for the case of friction) with the stiffness of the elastic loading portions. Since this elastic loading part ideally should be zero to represent the material continuity, the stiffnesses are chosen to be much higher than the Young's modulus of the material to minimize the displacements associated with the elastic loading. 

When the traction applied on the interface reaches the peak cohesion for tensile fractures $C_{I}^p$ or for shear fractures $C_{II}^p$, the cohesion starts weakening, and eventually it is totally broken, behaving as a secondarily activated fracture (Figure \ref{fig:app1}c). \add The friction curve follows linear slip-weakening law, originally proposed by \citet{ida1972a} and \citet{palmer1973}, which has been widely used for dynamic earthquake rupture modeling \citep[e.g.][]{andrews1976, aochi2002a, delapuente2009}. When the shear traction reaches to frictional strength $\tau_p$, it decreases down to the residual strength $\tau_r$ at critical slip distance for friction $\delta_{II}^{f,c} = D_c$ as shown in Figure \ref{fig:app1}d.
$\tau_p$ and $\tau_r$ are defined as 
\begin{equation}
\tau_p = f_s (-\sigma_n)
\label{eq:peakfric}
\end{equation}
\begin{equation}
\tau_r = f_d (-\sigma_n),
\label{eq:residualfric}
\end{equation}
where $\sigma_n$ is the normal stress on the contact interface. Note that the friction law is operating both on the main fault and the secondary fractures activated in the off-fault medium. \add[KO]{The residual traction on the fracture surface is zero for tensile fractures as long as the damage on the fracture surface is equal to one and as long as the fracture remains open, while the residual shear traction is kept at $\tau_r$ for shear fractures even after the shear cohesion is broken.}

The mixed mode fracture is evaluated by a damage parameter, $D$, which is defined as
\begin{equation}
D_i = \frac{\delta_i - \delta_i^{c,e}}{\delta_i^{c,c} - \delta_i^{c,e}} \quad i = I,II
\label{eq:damagedef1}
\end{equation}
\begin{equation}
D = \sqrt{D_I^2 + D_{II}^2} ~(0 \leq D \leq 1)
\label{eq:damagedef2}
\end{equation}
\begin{equation}
D^T = \frac{D_I}{D} = \left\{\begin{array}{ll}
        1, & \text{for purely tensile fracture}\\
        0, & \text{for purely shear fracture}
                \end{array}\right\},
\label{eq:damagetype}
\end{equation}
where $D_i$ ($i = I, II$) is the components of damage for tensile and shear fractures,  $\delta_i$ is normal and tangential displacement, $\delta_i^{c, e}$ is the initial critical displacement for elastic loading, $\delta_{i}^{c,c} - \delta_{i}^{c,e}$ is the maximum displacement during linear-softening, where $\delta_{i}^{c,c}$ is the initial critical displacement for linear-weakening part,  $D$ is the degree of damage and $D^T$ indicates the type of damage.
Similar expressions can be found in \citet{rougier2011} and \citet{lisjak2014}.

Since we used a linear softening law, the fracture energy associated with the cohesion for tensile (mode I) and shear (mode II) fractures $G_{IC/IIC}^c$ (i.e., the energy required to completely break the connection of the contact) is evaluated as
\begin{equation}
G_{iC}^c = \frac{1}{2} C^p_{i} \left ( \delta_{i}^{c,c} - \delta_{i}^{c,e} \right )  \quad i = I,II.
\label{eq:GIC}
\end{equation}
The fracture energy for friction $G_{IIC}^f$ is, following \citet{palmer1973}, described as
\begin{equation}
G_{IIC}^f = \dfrac{1}{2} D_c \left(\tau_p - \tau_r \right).
\label{eq:GIICf}
\end{equation}
Note that the elastic loading part for friction $\delta_{II}^{f,e}$ is much smaller than $D_c$, so that the representation of fracture energy $G_{IIC}^{f}$ by equation (\ref{eq:GIICf}) is acceptable even without the consideration of elastic loading part.

In this study, we assume that the fracture energy on the main fault is kept constant with depth\add{, denoted as $G_{IIC}^{f*}$}. Thus $D_c$ decreases with depth as follows:
\begin{equation}
D_c(z)= \frac{2G_{IIC}^{f*}}{\left( f_{s} - f_{d} \right) \left\{-\sigma^0_{yy}(z) \right\} }.
\label{eq:Dcwithdepth}
\end{equation}

\change[KO]{$\delta_{i}^{c,e}$ and $\delta_{i}^{c,c}$ ($i = I, II$) are derived with the stiffness of elastic loading part and given fracture energy,}{$\delta_{i}^{c,e}$ and $\delta_{i}^{c,c}$ ($i = I, II$) are derived with the stiffness of elastic loading part and given fracture energy. The shear fracture energy is estimated from the experiments and observations, following the scaling law between the fracture energy and the amount of slip} \citep{viesca2015, passelegue2016b}. \add[KO]{It provides a reasonable assumption of fracture energy on the fault and in the off-fault medium, corresponding to the mean slip on the main fault and on the off-fault fractures during simulation. 
The fracture energy on the main fault is assigned to be two orders of magnitude higher than that of individual off-fault fractures because the slip on the main fault is larger than that of the off-fault fractures when modeling with a single planar fault.} We assume $\delta_{i}^{c,e} = \delta_{i}^{f,e}$ so that the cohesion and the friction start weakening at the same amount of slip. The detailed formulations can be found in \citet[][Chapter 2]{okubo2018h}.

\subsection{Parametrization for peak cohesions}
To determine $C^p_{II}$, we used the closeness to failure $d_{MC}$, which indicates the safety of the initial stress state to the failure of the material represented by the ratio of the radius of the Mohr's circle to the distance to the Mohr-Coulomb criteria \citep[see also][]{templeton2008}.
Let $\sigma_1$ and $\sigma_2$ be the maximum and minimum compressive principal stresses. Then $d_{MC}$ is derived from geometrical relationships as
\begin{align}\nonumber
d_{MC} &= \frac{\sigma_2 - \sigma_1}{2C^p_{II} \cos{\phi} - (\sigma_1 + \sigma_2)} \\
&= \dfrac{\left( \dfrac{\sigma_1}{\sigma_2} - 1 \right)}  {\left( \dfrac{\sigma_1}{\sigma_2} + 1 \right) - 2\left(\dfrac{C^p_{II}}{\sigma_2} \cos{\phi} \right)},
\label{eq:dMC}
\end{align}
where $\phi$ is the friction angle as $\tan{\phi} = f_{s}$. $d_{MC}<1$ means no failure and $d_{MC}\ge1$ implies the initiation of failure in shear on the corresponding plane. Note that $d_{MC}$ locally changes due to perturbations of the stress field.

In the case study, initial $d_{MC}$ is kept constant with depth for the fair comparison between the different stress states. By assuming the constant orientation of maximum compressive principal stress $\Psi$ and seismic ratio $S$, the ratio of principal stresses ${\sigma_{1}}/{\sigma_{2}}$ is also kept constant with depth. Thus from equation (\ref{eq:dMC}), the ratio $C^p_{II}/\sigma_2 $ has to be kept constant to obtain an equal closeness to failure with depth, implying $C^p_{II}$ must increase linearly with depth. Therefore we first calculate $\sigma_{ij}^0$ as described in section \ref{sec:modeldescription}, and then we derive $C^p_{II}$ as
\begin{equation}
C^p_{II} = \frac{\sigma_2 - \sigma_1 + d_{MC} (\sigma_1 + \sigma_2) \sin{\phi}}{2 d_{MC} \cos{\phi}},
\label{eq:CFII}
\end{equation}
where $d_{MC}$ should be chosen carefully to avoid $C^p_{II}$ being negative. $C^p_{I}$ is chosen from the experiments \citep{cho2003}, kept constant with depth. We assume the acceptable range for $C^p_{I}$ is between 1-10 MPa. 

\subsection{Process zone size}
\label{sec:processzonesize}
The quasi-static process zone size $R_0$ is used to nondimensionalize length scale as it characterizes the scale of dynamic earthquake ruptures \citep{poliakov2002,rice2005}, which is described as

\begin{equation}
R_{0}(z) = \frac{9\pi}{16(1-\nu)} \frac{\mu G_{IIC}^{f}}{ \left[ (f_{s} - f_{d}) \left\{-\sigma^0_{yy}(z)\right\} \right] ^2},
\label{eq:processzonesizeconstGIIC}
\end{equation}
where $\nu$ is Poisson's ratio and $\mu$ is shear modulus. \remove{As shown in equations (\protect\ref{eq:processzonesizeconstGIIC}),} $R_{0}(z)$ decreases with depth as a function of $\left\{ -\sigma^0_{yy}(z) \right\} ^{-2}$. Since the size of potential failure area is of the same order of magnitude as $R_{0}(z)$ \add[KO]{in the analysis with steady-state crack} \citep[][]{poliakov2002}, \add[KO]{the damage zone size is also expected to decrease when assuming constant $G_{IIC}^f$ with depth. Although we model a spontaneous rupture propagation, the results of flower-like structure as shown in Figure \protect\ref{fig:damagezonewidth} can be explained by this estimation as the rupture velocity for the case of sub-Rayleigh rupture (S=1.0) with off-fault damage converges to slightly below of its limiting speed (Figure} \ref{fig:rupturevelocity}). 

\remove{, as mentioned by \protect \citet{rice2005}, which is verified by this study. }

The dynamic process zone size $R_{f}(v_r)$ is generally inversely proportional to the rupture velocity $v_r$, given by \citet[][eq. 6.16]{rice1980a} and \citet[][eq. 6.2.35]{freund1990}.  $R_{f}(v_r)$ gradually shrinks and asymptotically converges to zero as the rupture velocity approaches its limiting speed, which is known as Lorentz contraction.

\subsection{Potential failure area}
\label{app:potentialfailurearea}
We superimposed the stress concentration in Figure \ref{fig:fracmechanism} to highlight the potential failure area, where the secondary fractures are likely to be activated. For tensile fracture, we used the normalized first stress invariant 
\begin{equation}
\dfrac{I_1(t)}{I_1^{\text{init}}} = \dfrac{\sigma_{kk}(t)}{\sigma_{kk}^0},
\label{eq:I1}
\end{equation}
where $I_1^{\text{init}}=I_1(0)$ and $\sigma_{kk} = \sigma_{xx}+\sigma_{yy}+\sigma_{zz}$\add[KO]{, where $\sigma_{zz}=\nu(\sigma_{xx}+\sigma_{yy})$ in plane strain condition}. The small $I_1(t)/I_1^{\text{init}}$ thus indicates less confining pressures.
For shear fracture, we used the normalized closeness to failure $d_{MC}/d_{MC}^\text{init}$. The large $d_{MC}/d_{MC}^\text{init}$ indicates that the stress state is close to shear failure.

\section{Cross-validation of 2-D FDEM for earthquake rupture modeling}
\label{sec:appb}
We performed cross-validation of the FDEM-based software tool, HOSSedu (denoted as HOSS in \change{the following}{this} section), to assess the achievable accuracy of dynamic earthquake rupture modeling with purely elastic medium, i.e. no off-fault damage, by comparing the results with HOSS to those with other numerical schemes. We chose the finite \change{different}{difference} method (FDM), the spectral element method (SEM) and the boundary integral equation method (BIEM) as comparison basis as they have been verified in previous studies \citep[e.g.][]{koller1992, day2005, kaneko2008}. 

The cross-validation effort for HOSS is based on a similar process to \citet{kaneko2008}. The first arrival time of the rupture is \change{a suitable benchmark}{used} to evaluate the numerical precision of the rupture solution \citep{day2005}. 
In this study, the rupture arrival time is defined \remove{at the time }when the shear traction reaches the peak strength $\tau_p$.
We followed the version 3 of the benchmark problem proposed by the Southern California Earthquake Center/U.S. Geological Survey (SCEC/USGS) dynamic earthquake rupture code verification exercise \citep{harris2009}, commonly used for cross-validating numerical schemes \citep{day2005,kaneko2008,rojas2008,delapuente2009}. 
The model is originally described in 3-D so that the 2-D analog model was used in this study, similar to \citet{rojas2008}, \citet{kaneko2008}, and \citet{delapuente2009}. 

Figure \ref{fig:cv1}a shows the the comparison of slip velocity history at x = 9 km from the center of the main fault. The results of HOSS are compared to FDM, SEM and BIEM, where the grid spacing on the fault $\Delta s$ is chosen for the highest resolution as $\Delta s = 8$ m ($R_{0}/\Delta s = 116$) for HOSS, FDM and BIEM and $\Delta s = 10$ m ($R_{0}/\Delta s = 93$) for SEM. The slip velocity history of HOSS is consistent with the other numerical schemes except for the peak slip velocity. The peak slip velocity of HOSS is 4.1 percent smaller than that of BIEM and the rupture arrival time is slightly faster than the others. Both of the small discrepancies are explained by the artificial viscous damping. There is no viscous damping for BIEM and FDM, whereas the Kelvin-Voigt viscous damping is used for SEM, and the Munjiza viscosity is used for HOSS. Although the viscous damping causes small reduction of the peak velocity and shortens the rupture arrival time, the high-frequency numerical noise is significantly removed for the result with HOSS.  
It is notable that the comparison of HOSS to BIEM is no longer fair due to the artificial viscous damping, so that the evaluation of the effect of viscous damping on the rupture propagation is worthwhile, discussed later in this section.

Figure \ref{fig:cv1}b shows the grid convergence of HOSS and the others.
The numerical accuracy as a function of grid resolution is evaluated by the root-mean-square (RMS) difference.
The RMS error of the rupture arrival time is calculated by the comparison to the benchmark solution provided by BIEM with highest resolution as it is semi-analytical solution. Although the RMS error is slightly higher than the FDM and SEM due to the viscosity, the convergence rate of HOSS is similar to BIEM, following the power law with the scaling exponent of 1.6 for HOSS and 1.4 for BIEM. Thus the numerical accuracy is assured with  appropriate $\Delta s$ for the required error range of earthquake rupture modeling.

Figures \ref{fig:cv1}c and \ref{fig:cv1}d show the RMS error of the rupture arrival time with various viscous values, grid resolutions and the number of points per edge. The circles indicate the examined combinations of viscosity and grid resolution, where the size of circles with monochromatic gradation represents the proportion of the viscosity to the theoretically derived critical viscosity (see also \citet{okubo2018h}).
The saddle of the RMS error around $\eta/M_v \Delta t = 10^2$, where $\eta$ is viscosity, $M_v$ is the Munjiza constant and $\Delta$t is time step, is explained by the competition between the numerical oscillation and the overdamped system. The convergence of the RMS error is better with two integral points per edge. Hence the grid resolution, viscosity and the number of points per face should be carefully chosen for the required numerical accuracy. Since the number of points per edge \change{should}{has to} be more than two \add[KO]{in order} to allow for the secondary fractures in the off-fault medium due to numerical reasons, we chose the appropriate grid size and viscosity from Figure \ref{fig:cv1}d for the case study with depth.

\section{Mesh dependency}
\label{sec:appc}
We examined two types of mesh to investigate mesh dependency associated with the coseismic off-fault damage. We made mesh \#1, which is used for the case study in main section, and mesh \#2, where the grid size on the fault is 5\% smaller than the mesh \#1 to change the mesh topology. We conducted the dynamic rupture simulation at 2 km depth with S=1.0 using these meshes. Figure \ref{fig:meshdepence_fracpattern}a shows the trace of off-fault fractures with each mesh. The damage zone width is consistent between them, whereas the damage pattern varies due to the different arrangement of potential failure planes in the off-fault medium. Small perturbation in the mesh topology thus changes the detailed damage pattern because of its chaotic aspects of the system.

However, statistical quantities are not influenced by the mesh topology. Figure \ref{fig:meshdepence_fracpattern}b shows the rose diagram of the orientation of off-fault fractures, which is in agreement between mesh \#1 and mesh \#2. In addition, the fraction of fracture type is also compatible between them. \change[KO]{It is important to average the orientation of potential failure plane as shown in the histogram to realize the independence from the mesh topology.}{To realize the independence from the mesh topology, the orientation of potential failure plane is uniformly averaged as shown in the histogram.}

Figure \ref{fig:meshdepence_radiation}a shows the spatial distribution of critical frequency and spectra associated with those meshes. The spatial distribution varies as it depends on the damage pattern, whereas both spectra show the enhanced high-frequency radiation regardless of mesh topology. Figure \ref{fig:meshdepence_radiation}b shows the comparison in the seismic efficiency $\eta_r$ and the contribution of fracture energy in the off-fault medium \add{$E_G^{\text{off}}/E_G^{\text{on}}$} as a function of time. $\eta_r$ is well consistent between meshes. Since we indirectly evaluate the $E_G^{\text{off}}$ with uncertainty of the energy dissipation by numerical viscous damping, we showed the estimation of $E_G^{\text{off}}/E_G^{\text{on}}$ with error bands. Both results are fairly overlapped, and show the increase in $E_G^{\text{off}}/E_G^{\text{on}}$ with rupture propagation, which is sufficient for the argument in section \ref{sec:overallenergybudget}.

In summary, although the detailed damage pattern depends on the mesh topology, the statistical quantities such as orientation of fractures, radiation and overall energy budget are not so much influenced by the mesh topology.
Furthermore, when considering geometrical complexity of the fault system, \change{the damage pattern is better determined due to the stress concentration caused by the fault geometry, which activates the off-fault fractures in dominant orientations.}{the damage pattern is dominantly determined by the regional stress and fault geometry due to the stress concentration caused by the geometrical complexity such as fault kinks or fault roughness.}

\acknowledgments

The cross-validation \add[KO]{of HOSSedu} was performed using the softwares of dynamic earthquake rupture modeling; MultiDimensional Spectral Boundary Integral code (MDSB), available at \url{https://pangea.stanford.edu/~edunham/codes/codes.html}, and Spectral Element Method tool for 2D wave propagation and earthquake source dynamics (SEM2DPACK), available at \url{http://web.gps.caltech.edu/~ampuero/software.html}. The mesh discretization was performed with the mesh generation toolkit, Trelis Pro.
\add[KO]{We acknowledge Los Alamos National Laboratory (LANL) Institutional Computing program for the computing resources provided for this work. Part of numerical simulation was also performed on the S-CAPAD platform, IPGP.
This work is supported by the LANL LDRD Program (\#20170004DR) and the PhD funding from Universit\'{e} Sorbonne Paris Cit\'{e} (USPC).}

\remove{The authors also thank Los Alamos National Laboratory (LANL) for the license of HOSSedu.}

\bibliography{./MasterBibliography.bib}
\clearpage

\renewcommand\thefigure{\arabic{figure}}    
\setcounter{figure}{0}    
\begin{figure}
\center
\noindent\includegraphics[width=0.85\textwidth]{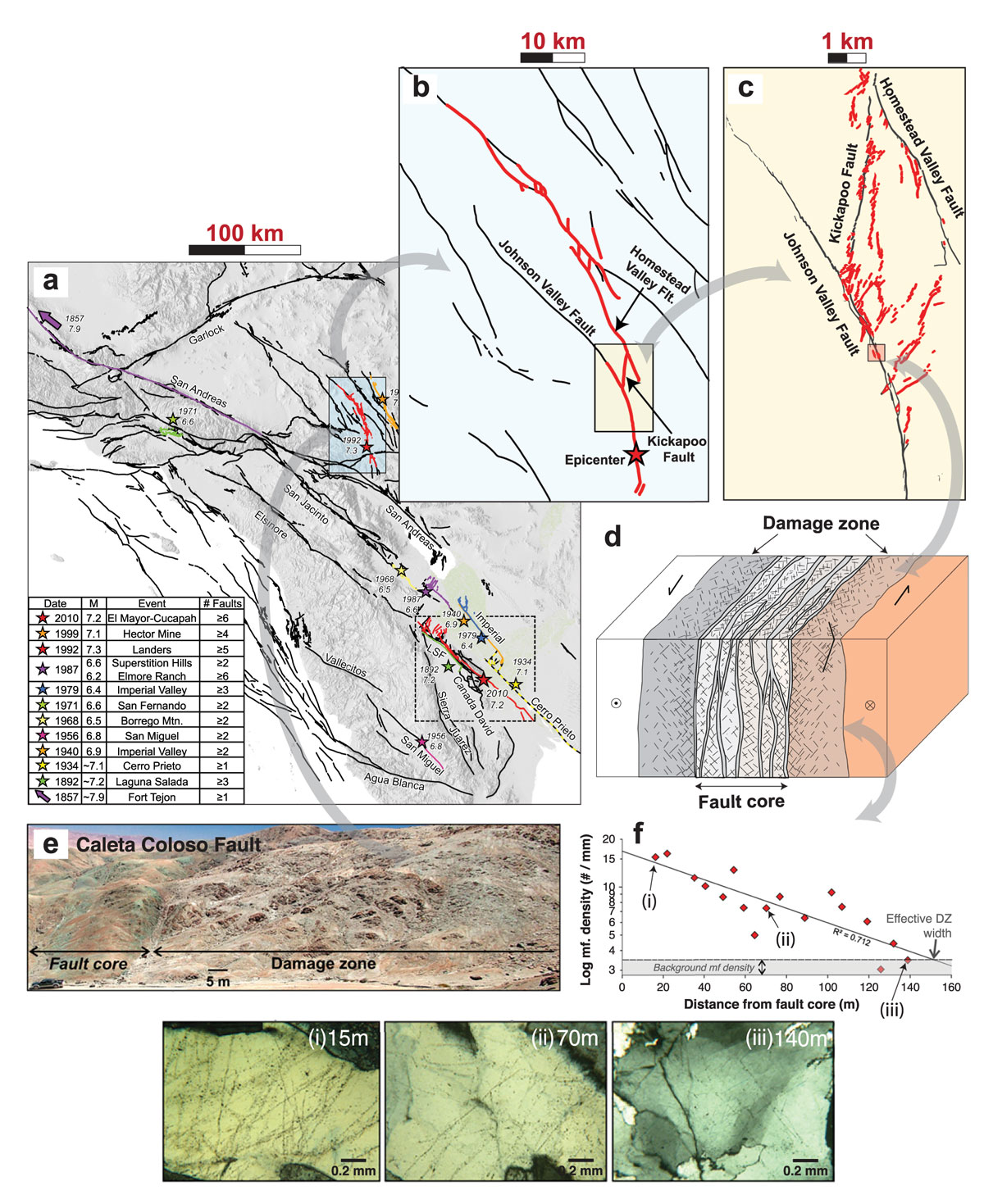}
\caption{Hierarchical structure of fault systems in a wide range of length scales. (a) Fault map of southern California \citep{fletcher2014}. Black lines indicate the fault traces. Stars and color lines indicate the epicenters and the rupture traces of historic earthquake events, respectively. (b) Fault map and the rupture traces (in red) associated with the 1992 Landers earthquake \citep[modified from][]{sowers1994}. (c) Smaller scale off-fault fracture network  \citep{sowers1994}.  (d) Schematic of fault zone structure, showing a fault core surrounded by damage zones \citep{mitchell2009}. (e, f) Fault damage zone of Caleta Coloso fault, the variation in microfracture (mf.) density within the damage zone as a function of distance from fault core, and pictures of microfractures at different distances from the fault core.  \citep{mitchell2012a}.}
\label{fig:hierarchicalfaultsystem}
\end{figure}

\begin{figure}
\center
\noindent\includegraphics[width=\textwidth]{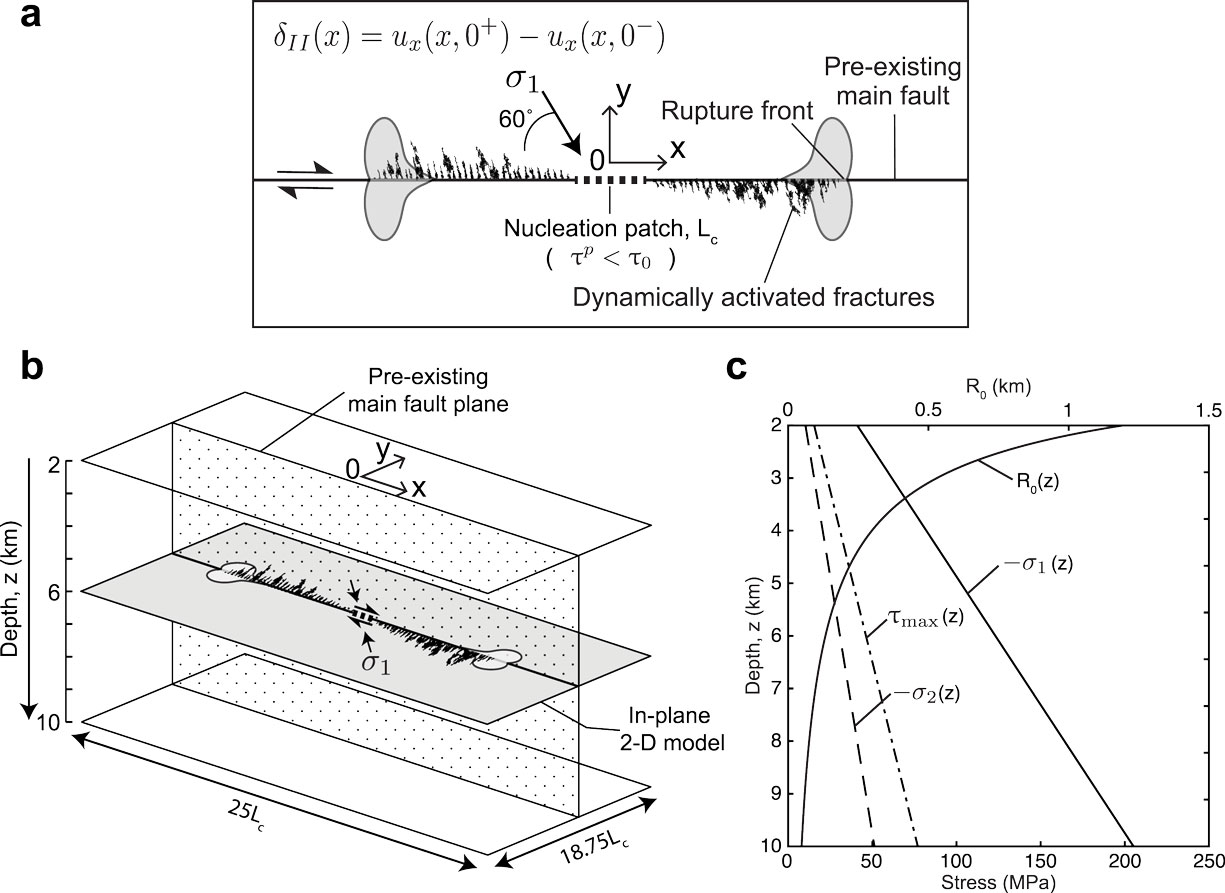}
\caption{Model description for the case study with depth. (a) 2-D strike-slip fault for dynamic rupture modeling with coseismic off-fault damage. The pre-existing fault is defined as the interface without cohesion. The orientation of maximum compressional principal stress $\sigma_1$ is fixed to 60$^\circ$ from the main fault. The slip on the fault $\delta_{II}$ is defined as the relative displacement.  (b) Schematic of case study with depth. (c) The evolution of initial stress state and quasi-static process zone size $R_0(z)$ with depth. $-\sigma_1(z)$, $-\sigma_2(z)$, $\uptau_{\text{max}}(z)$ indicate maximum principal stress, minimum principal stress and maximum shear traction, respectively.}
\label{fig:modeldescription}
\end{figure}

\begin{figure}
\center
\noindent\includegraphics[width=\textwidth]{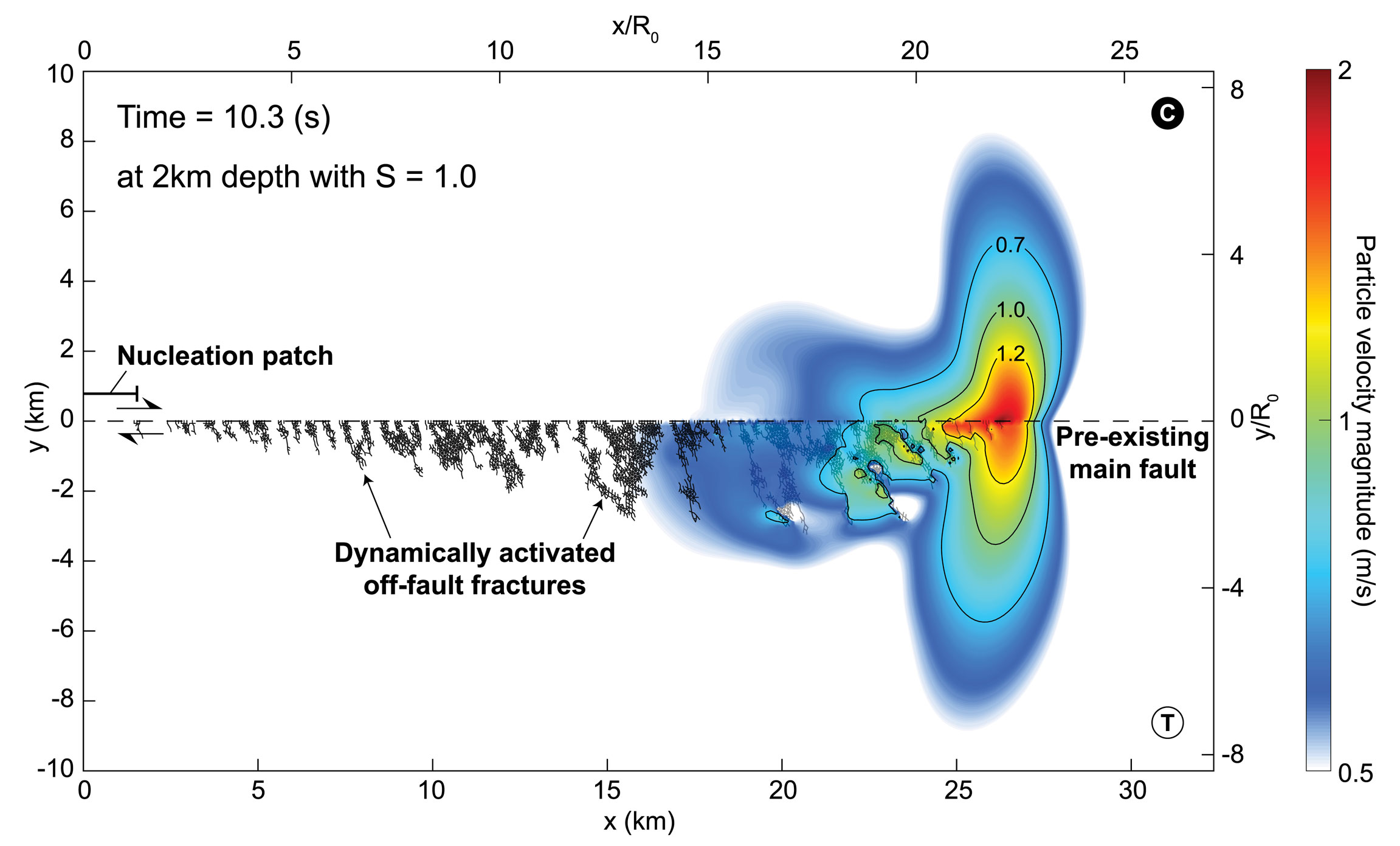}
\caption{Snapshot of the dynamic earthquake rupture with coseismic off-fault damage. We plot only the right part ($x$ $>$ $0$) as the result is symmetrical with respect to the origin. The initial stress state and the strength of material correspond to 2km depth with S=1.0. Color contour indicates the particle velocity magnitude. Dotted line indicates the main fault and the solid lines indicate the secondarily activated off-fault fractures. The bottom and left axis show the physical length scale, while the top and right axis show the nondimensionalized length scaled by $R_0$. "C" and "T" at right corners indicate compressional and extensional side of the main fault, respectively.}
\label{fig:snapshotSR}
\end{figure}

\begin{figure}
\center
\noindent\includegraphics[width=\textwidth]{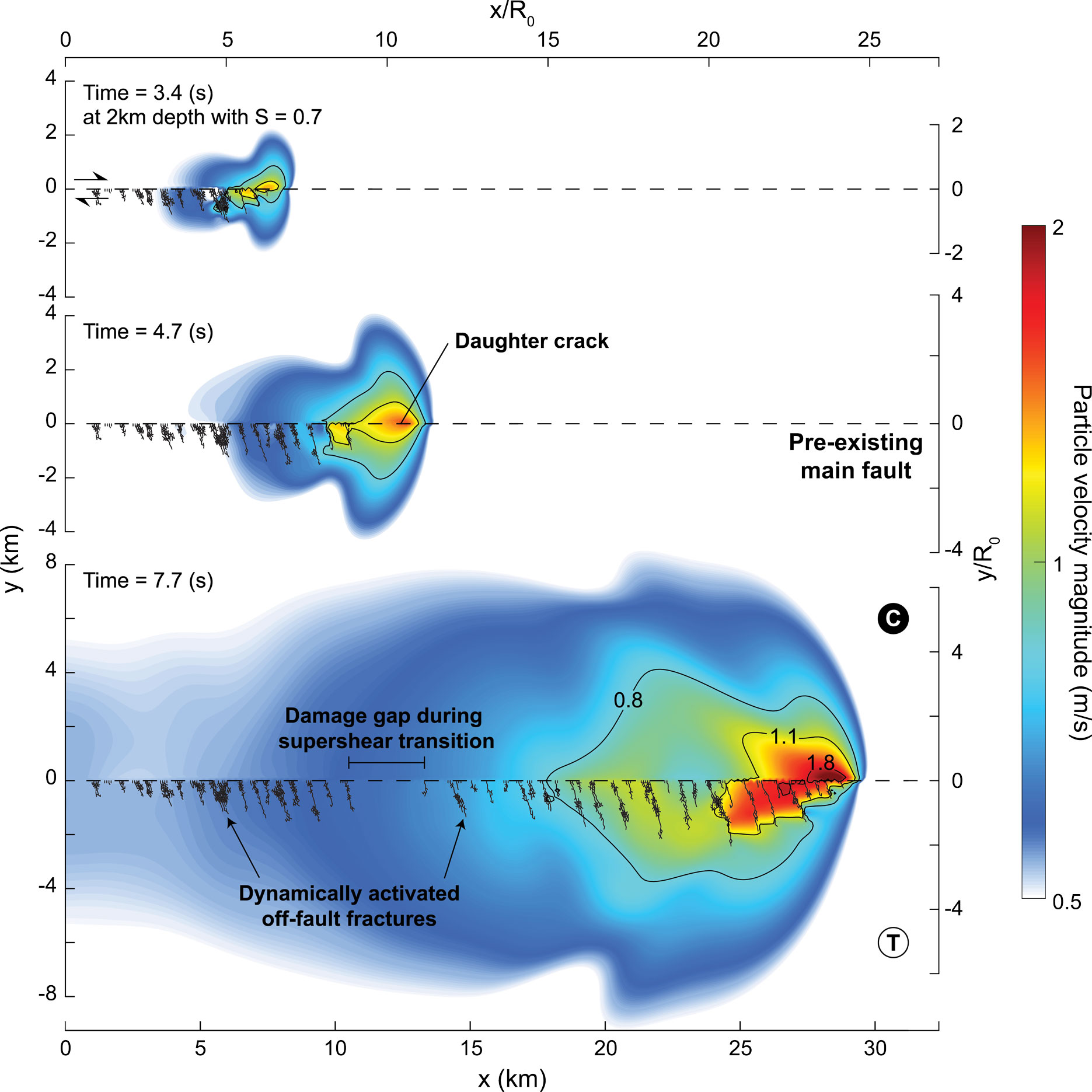}
\caption{Snapshots of supershear rupture at 2km depth with S=0.7. Color contour and lines indicate the same as Figure \ref{fig:snapshotSR}. The rupture velocity is sub-Rayleigh until T=3.4 s (top), then a daughter crack is born ahead of the sub-Rayleigh rupture front at T=4.7 s (middle), which transitions to the supershear (bottom). }
\label{fig:snapshotSS}
\end{figure}

\begin{figure}
\center
\noindent\includegraphics[width=\textwidth]{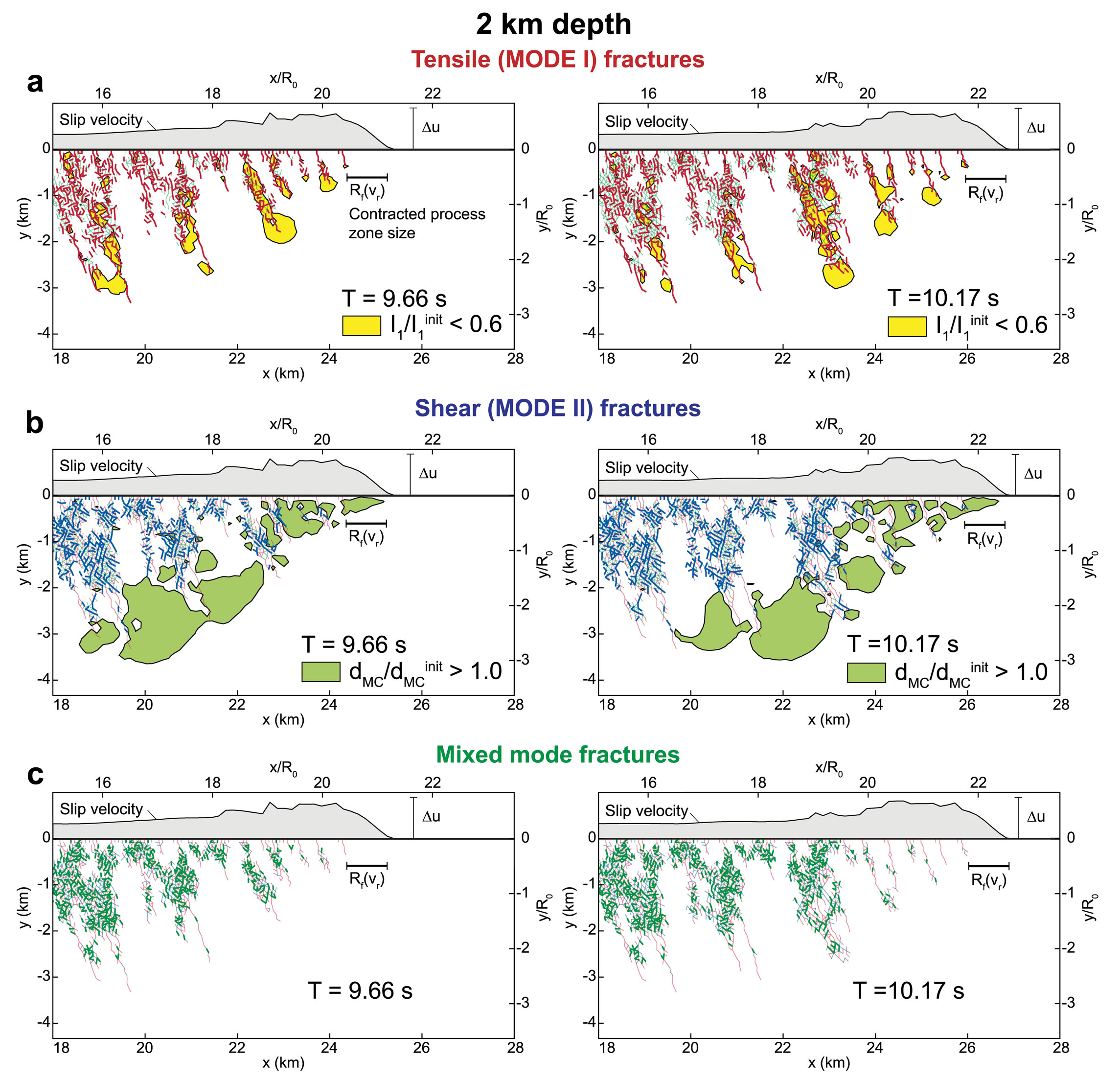}
\caption{Off-fault fracturing process in tensile, shear and mixed mode with S=1.0 at 2km depth. (a) Trace of tensile fractures at T=9.66s and 10.17s. Red heavy lines indicate the tensile fracture with damage type $D_T$ $\geq$ $0.9$  (eq. \ref{eq:damagetype}) and damage $D$ $\geq$ $0.01$. Solid line on the top of the main fault indicates the slip velocity on the main fault. The filled area in yellow shows the potential failure area where the ratio of the first stress invariant to its initial value $I_1(t)/I_{1}^{init}$ is less than 0.6. The lighter lines in the fracture network indicate shear and mixed mode fractures. $R_f(v_r)$ shows the dynamic process zone size of the earthquake rupture on the main fault. (b) Trace of shear fractures. Blue heavy lines indicate the shear fracture with damage type $D_T$ $\leq$ $0.1$ and damage $D$ $\geq$ $0.01$. The filled area in green shows where the ratio of closeness to failure to its initial value $d_{MC}/d_{MC}^{init}$ $>$ $1.0$. $d_{MC}^{init}$ is uniformly equal to 0.4 in the domain. (c) Trace of mixed fractures with  $0.1$ $<$ $D_T$ $<$ $0.9$ and $D$ $\geq$ $0.01$.}
\label{fig:fracmechanism}
\end{figure} 

\begin{figure}
\center
\noindent\includegraphics[width=0.75\textwidth]{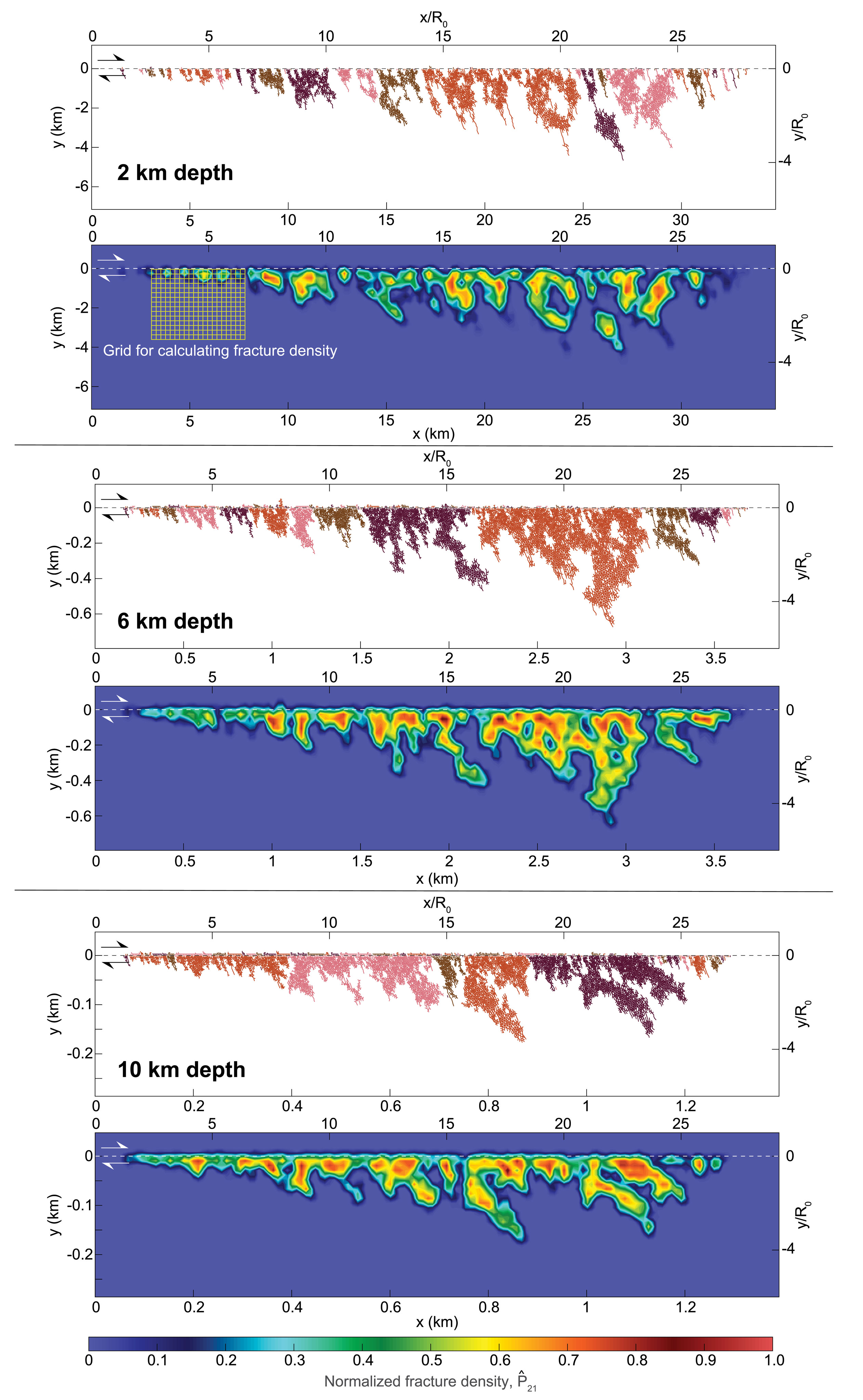}
\caption{Off-fault fracture network and spatial distribution of fracture density with depth. The results are at the final snapshot of simulations with S=1.0. An isolated fracture network, in which all small fractures connect with each other, is indicated by different colors. Color contour indicates the normalized fracture density $\hat{P}_{21}$.}
\label{fig:fracmechwithdepth}
\end{figure} 

\begin{figure}
\center
\noindent\includegraphics[width=\textwidth]{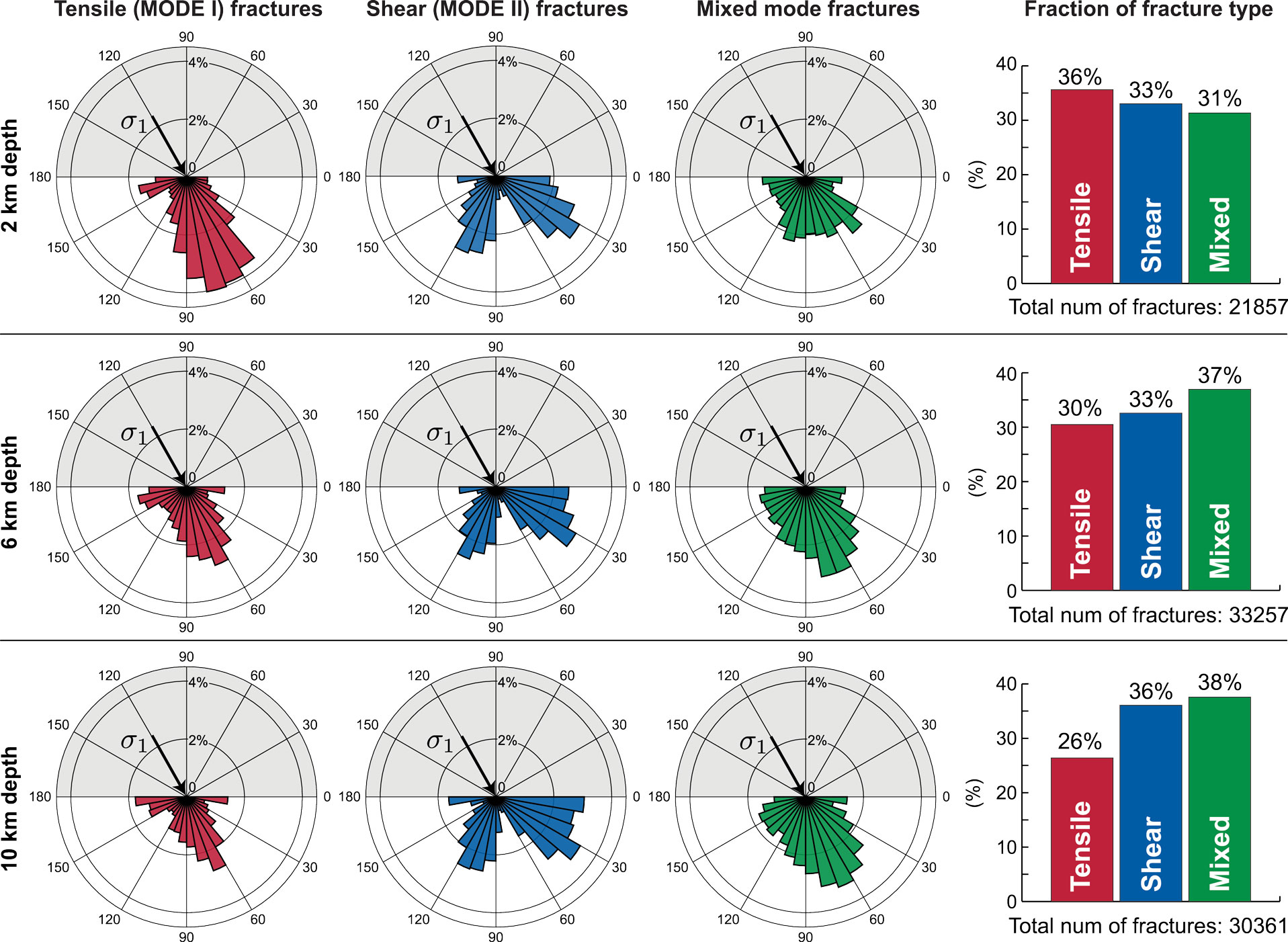}
\caption{Rose diagram showing the orientation of secondarily activated fractures obtained from Figure \ref{fig:fracmechwithdepth} and the fraction of each type of fracture. Size of bars in rose diagram is normalized by the sum of all types of fracture. The arrow indicates the orientation of maximum compressive principal stress $\sigma_1$ on the main fault.}
\label{fig:rosediagram}
\end{figure} 

\begin{figure}
\center
\noindent\includegraphics[width=0.6\textwidth]{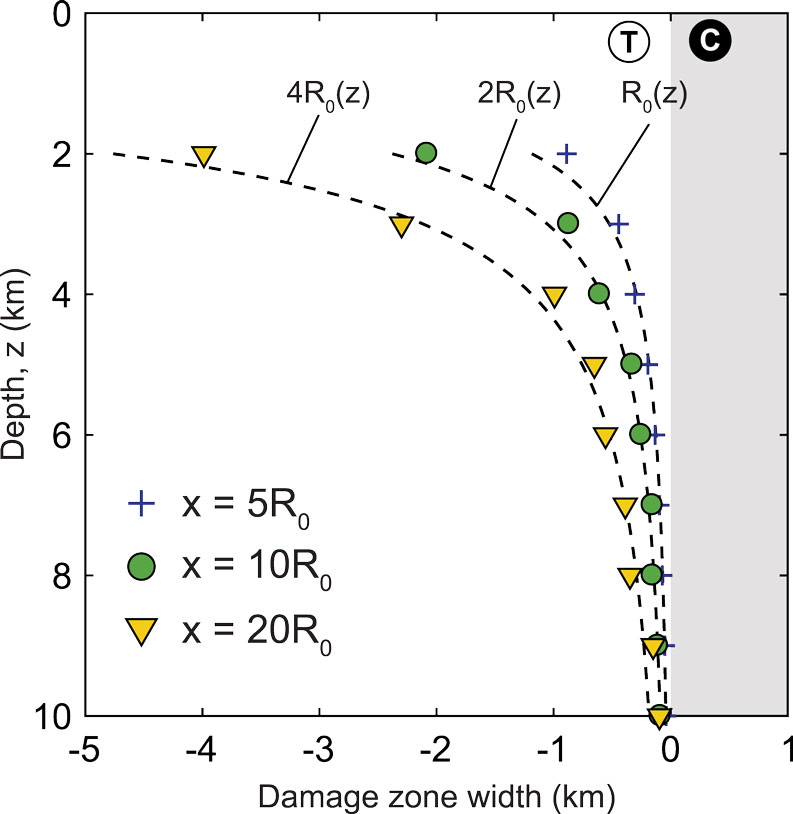}
\caption{Evolution of damage zone width with depth. Markers indicate the damage zone width obtained from the case study. Type of markers indicate the position on the main fault at which the damage zone width is evaluated. The dotted lines indicate the quasi-static process zone size scaled by constant factors.}
\label{fig:damagezonewidth}
\end{figure}

\begin{figure}
\center
\noindent\includegraphics[width=\textwidth]{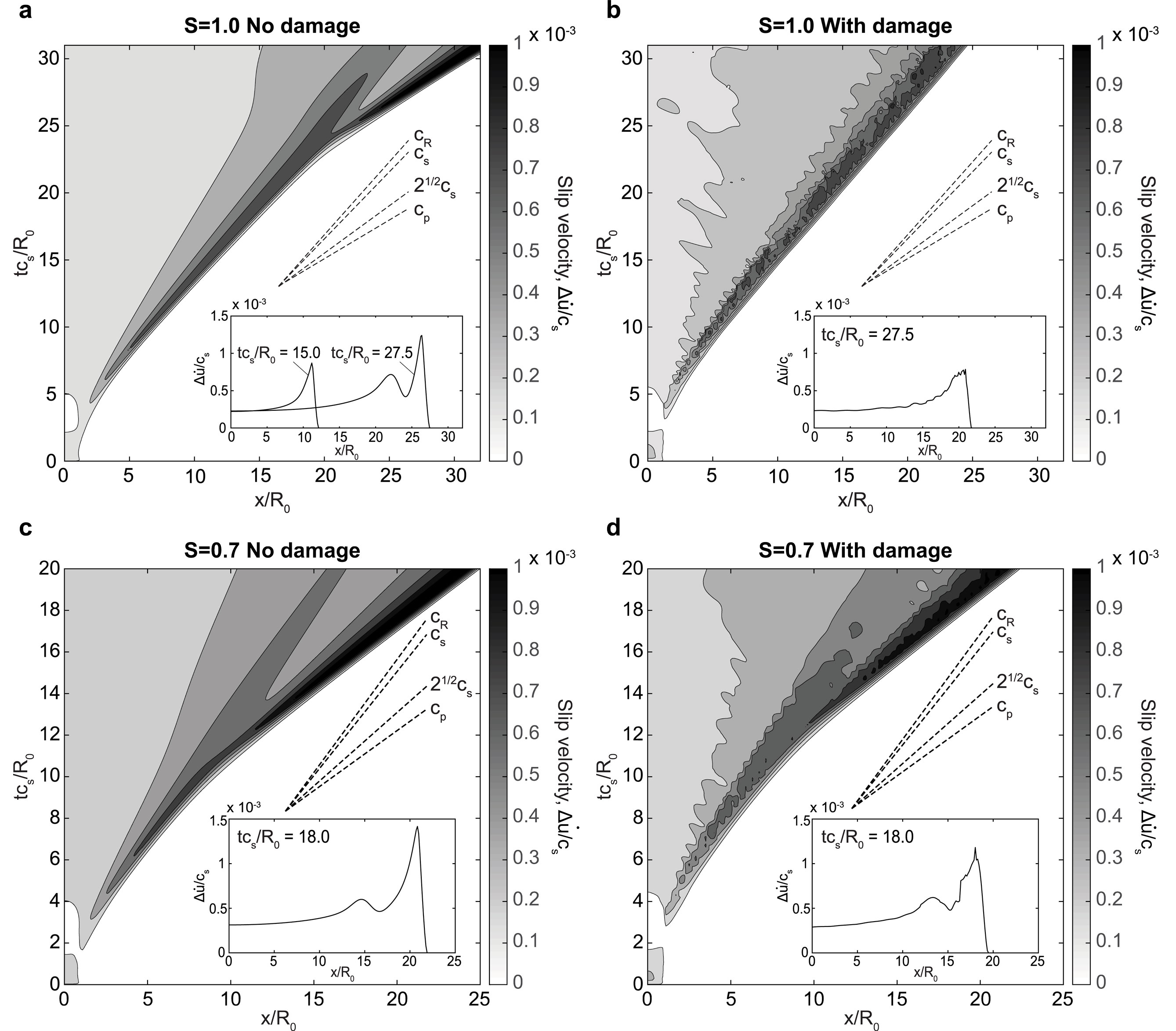}
\caption{The evolution of slip velocity in time and space at 2km depth. There are four cases: (a) S = 1.0 with no damage in the off-fault medium (b) S = 1.0 with damage (c) S = 0.7 with no damage (d) S = 0.7 with damage. For the cases without damage, we set extremely high cohesion for both tensile and shear fractures so that the off-fault medium behaves as a purely elastic material. 
Color contour indicates the slip velocity. Dotted lines indicate the reference of the slope corresponding to each wave velocity. Insets show the distribution of slip velocity on the main fault at certain time.}
\label{fig:xtplot}
\end{figure} 

\begin{figure}
\center
\noindent\includegraphics[width=0.8\textwidth]{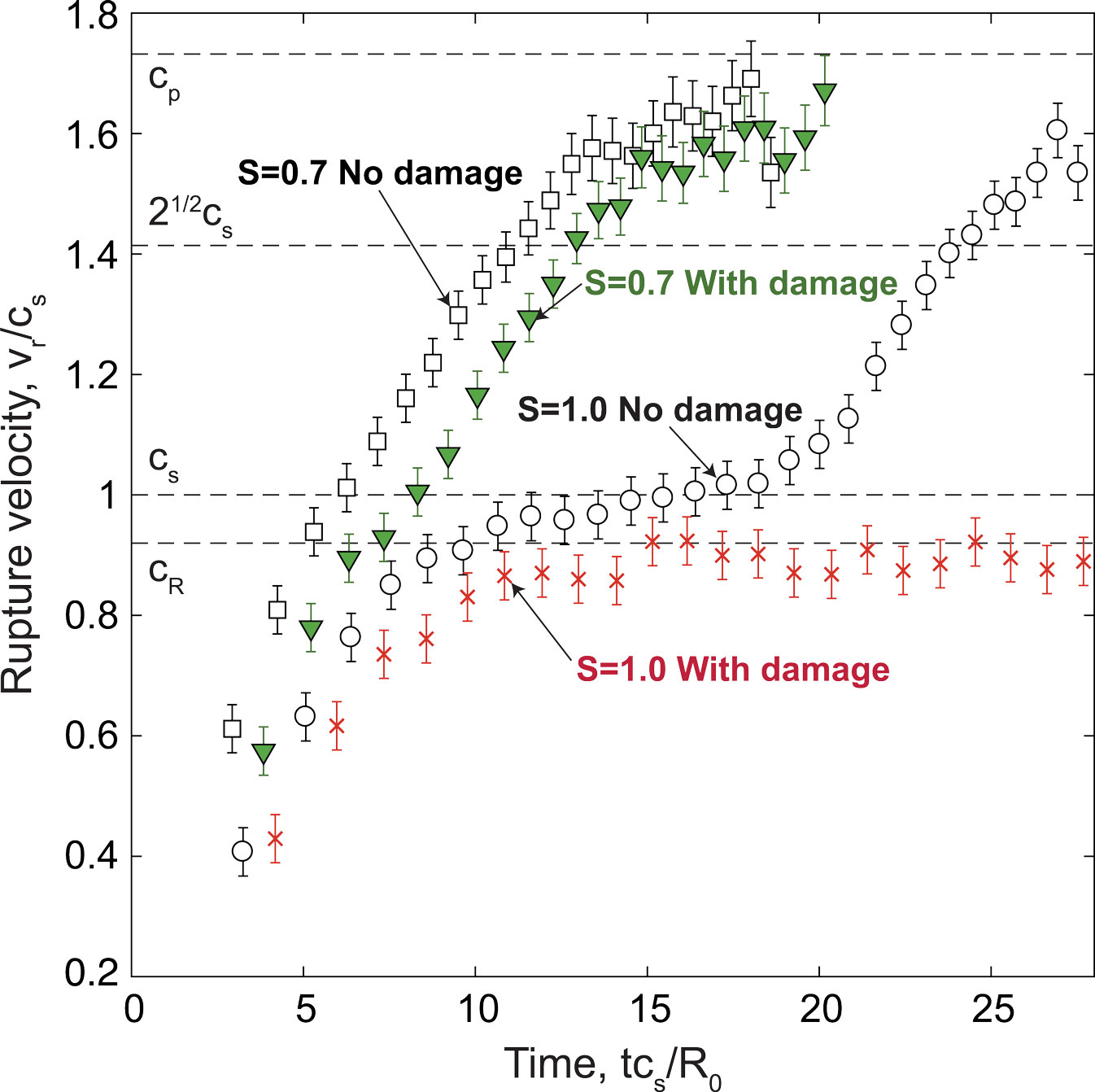}
\caption{Rupture velocity inferred from Figure \ref{fig:xtplot}. Due to inherent discretization errors, it is difficult to precisely capture the jump of rupture velocity from sub-Rayleigh to supershear. The error is estimated from the difference between the slope of $c_R$ and $c_s$, the grid spacing and the sampling rate of slip velocity.}
\label{fig:rupturevelocity}
\end{figure} 

\begin{figure}
\center
\noindent\includegraphics[width=0.75\textwidth]{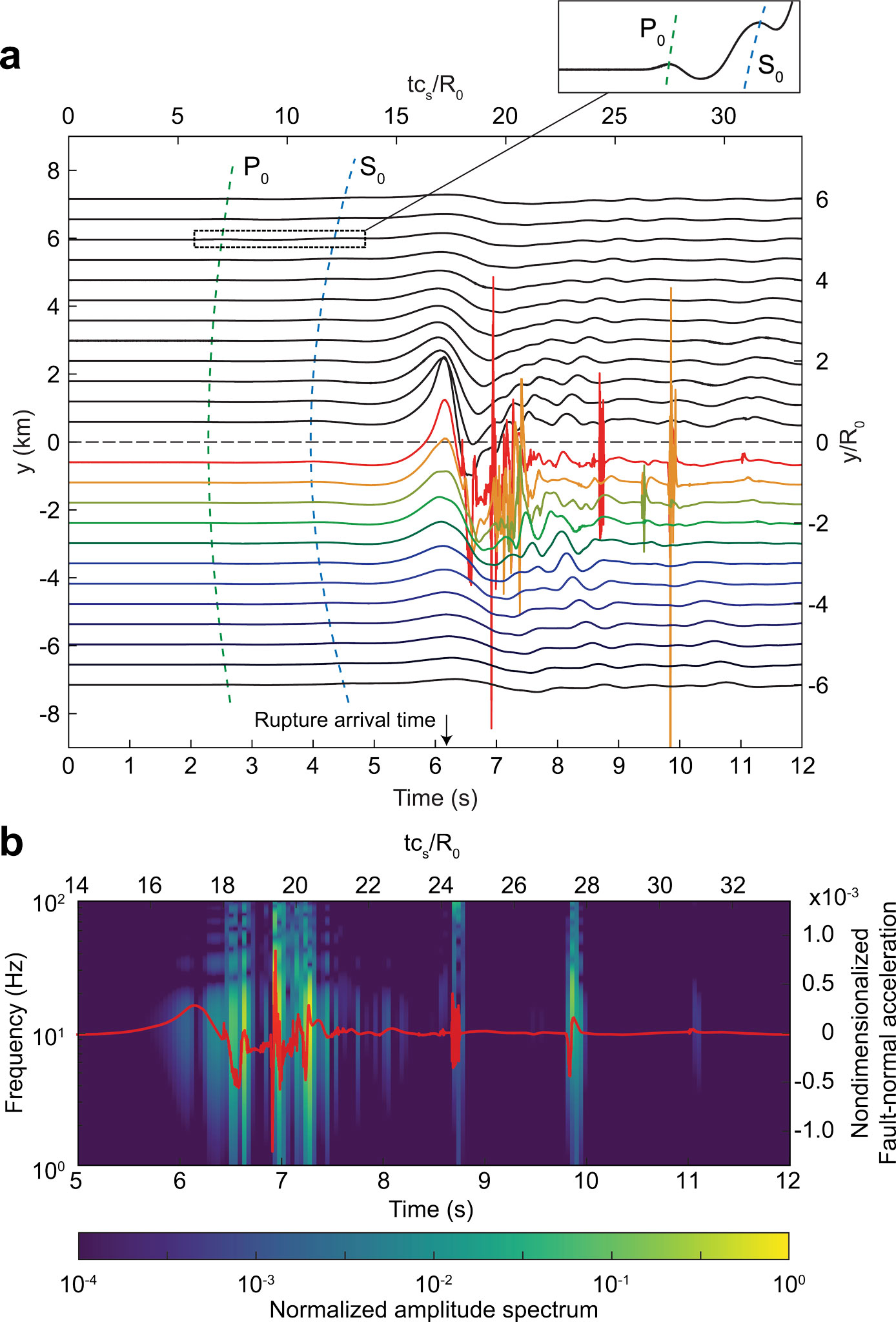}
\caption{(a) Fault-normal acceleration at $x$$=$$12.4R_0$, 2km depth with $S$$=$$1.0$. Line colors of waveform indicate the fault-normal distance. Dotted lines indicate the theoretical P and S wave arrival time. The inset shows magnified signals, where the P and S arrival can be found. The rupture arrival time at the location of stations ($x$ $=$ $12.4R_0$) is 6.2s, indicated by the arrow. (b) Spectrogram of the near-field ground acceleration ($y$$=$$-0.5R_0$). The amplitude \add[KO]{spectrum} is normalized by its maximum value over the time.}
\label{fig:waveform}
\end{figure} 

\begin{figure}
\center
\noindent\includegraphics[width=0.75\textwidth]{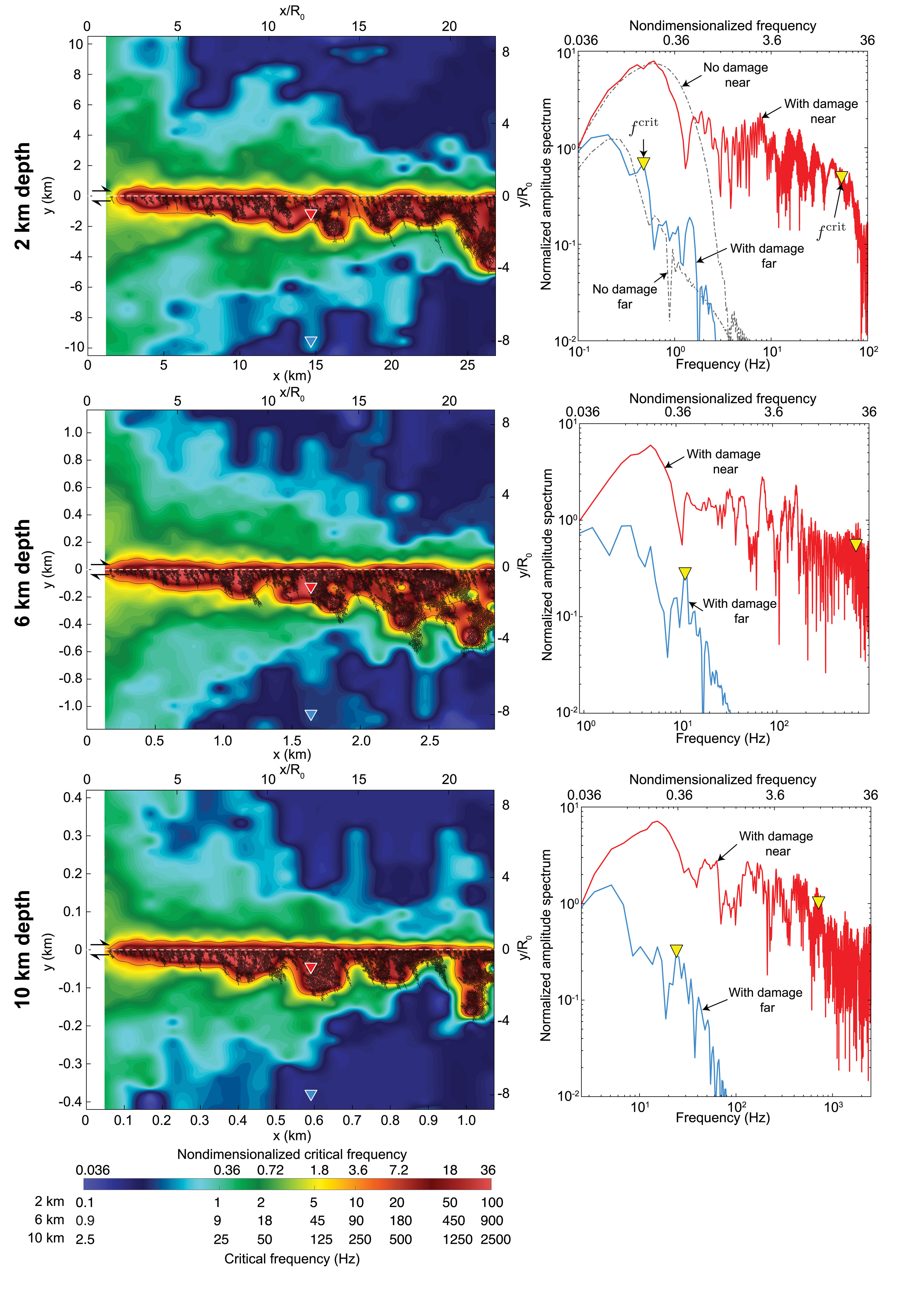}
\caption{Spatial distribution of critical frequency and spectra of fault-normal acceleration with depth. Color contour shows the critical frequency. The off-fault fractures are superimposed with the black lines. Inverted triangles indicate the locations of spectra.  The spectra for \change[KO]{the no damage case}{the cases without off-fault damage} at 2km depth are indicated by the \change{dotted}{dashed} lines in gray.}
\label{fig:freqwithdepth}
\end{figure} 

\begin{figure}
\center
\noindent\includegraphics[width=0.8\textwidth]{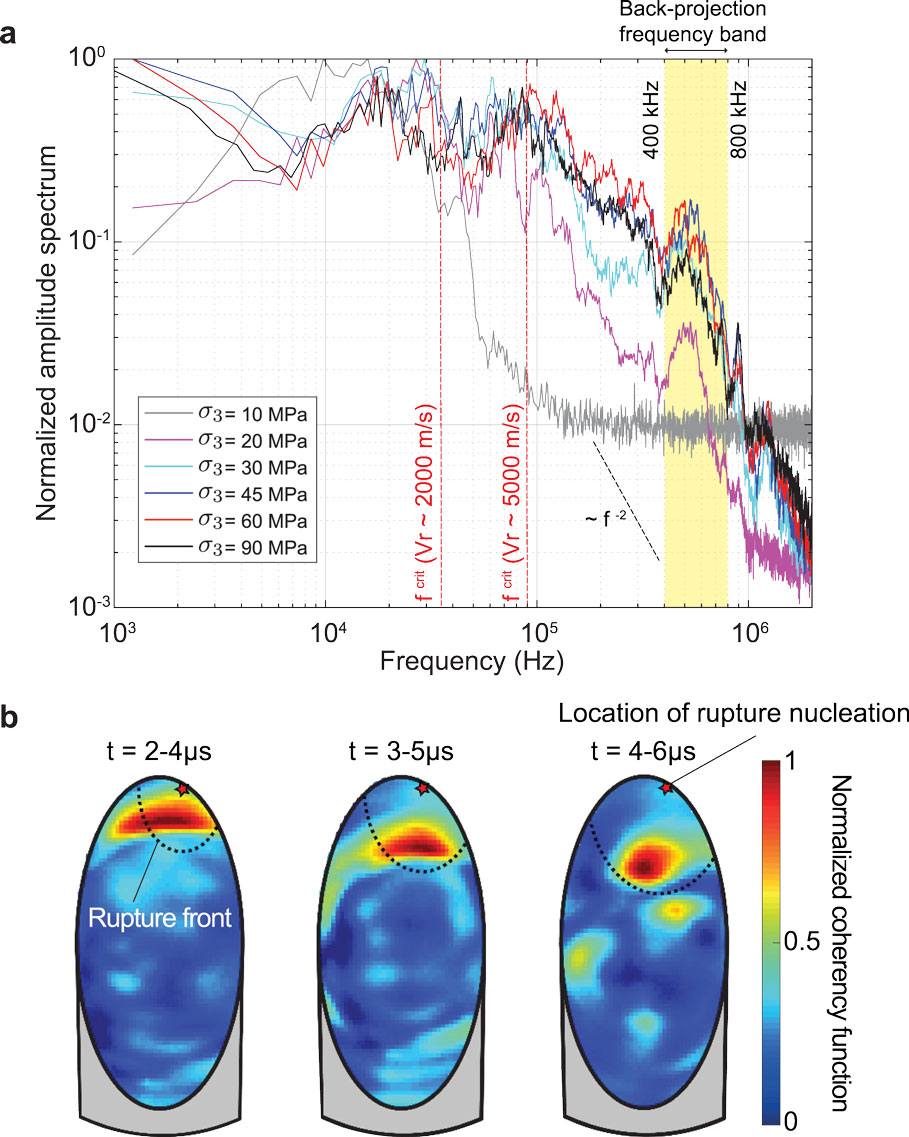}
\caption{Enhanced high-frequency radiation and back-projection analysis in laboratory experiments \citep{marty2019}. (a) Fourier spectra with different confining pressures. Red dashed lines indicate the theoretical critical frequencies at $v_r$ $=$ $2000$m/s and $v_r$ $=$ $5000$m/s. Highlighted box indicates the frequency band used for the back-projection analysis. (b) Snapshots of back-projection results (bandpass filtered from 400kHz to 800 kHz) with $\sigma_3$ $=$ $90$MPa at t= 2-4$\mu$s, 3-5$\mu$s and 4-6$\mu$s relative to the onset of rupture. Red star indicates the nucleation position. Dashed line indicates the theoretical rupture front. Color contour shows the normalized coherency function, which indicates the most likely location of the origin of signals within the frequency band.}
\label{fig:experiments}
\end{figure}

\begin{figure}
\center
\noindent\includegraphics[width=1.0\textwidth]{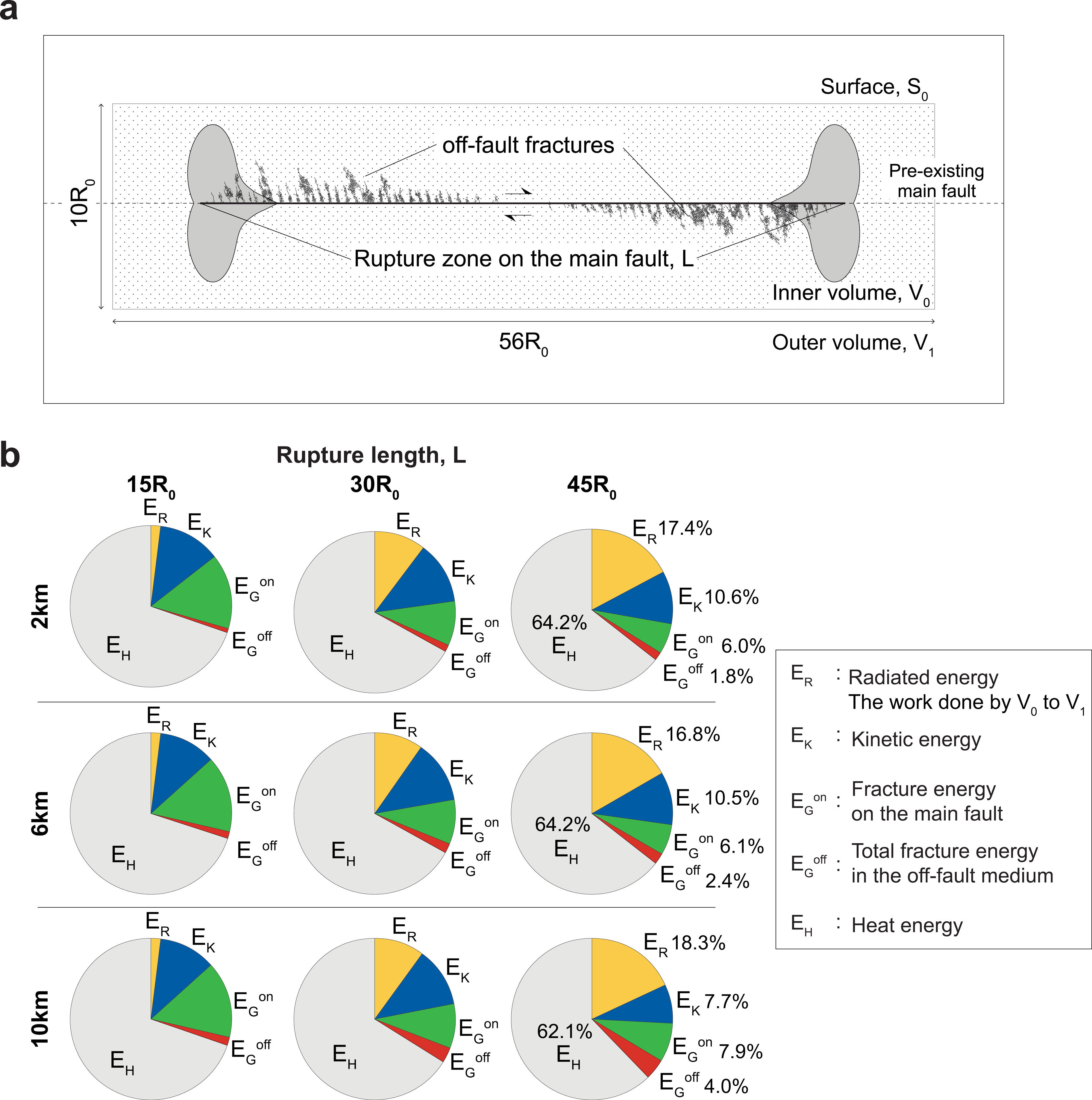}
\caption{Schematic of overall energy budget and the fraction of energy components. (a) Schematic of overall energy budget. The dotted area shows the inner volume $V_0$ with surface $S_0$, where the energy budget is evaluated. The inner volume is rectangular with unit thickness in our calculation. The size of the target area is arbitrary chosen as 10$R_0$ $\times$ 56$R_0$. (b) Fraction of energy components against $-(\Delta W + E_{S_0}^0)$ as a function of rupture length with depth. The results are for the case with $S=1.0$.}
\label{fig:energyschematic}
\end{figure}

\begin{figure}
\center
\noindent\includegraphics[width=1.0\textwidth]{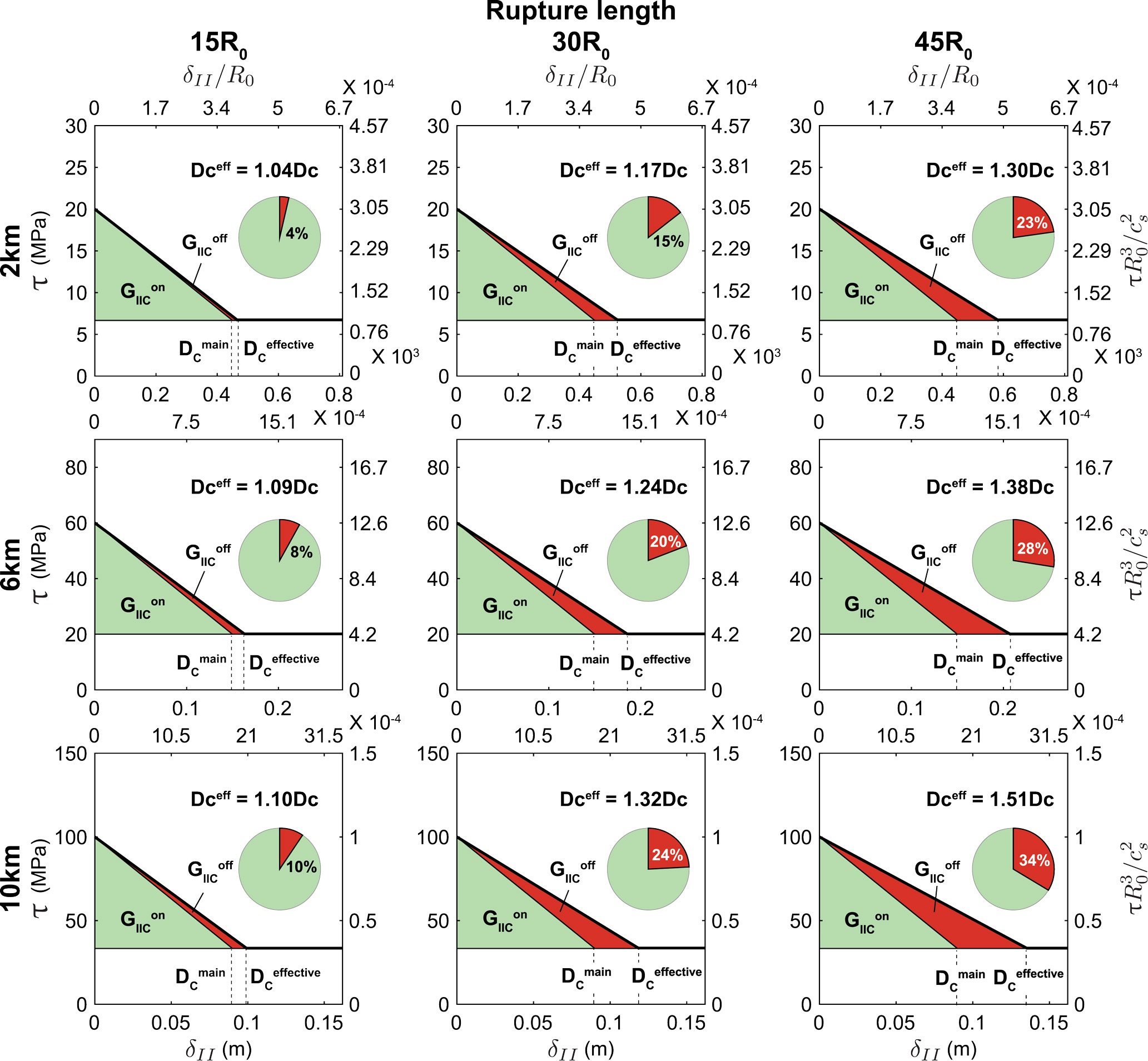}
\caption{Effective $D_c$ derived from Figure \ref{fig:energyschematic}b. $G_{IIC}^{\text{off}}$ indicates the total fracture energy dissipated due to the coseismic off-fault damage.The pie chart indicates the fraction of $E_G^{\text{off}}$ against $E_G^{\text{on}}$.}
\label{fig:effectiveDc}
\end{figure} 

\begin{figure}
\center
\noindent\includegraphics[width=1.0\textwidth]{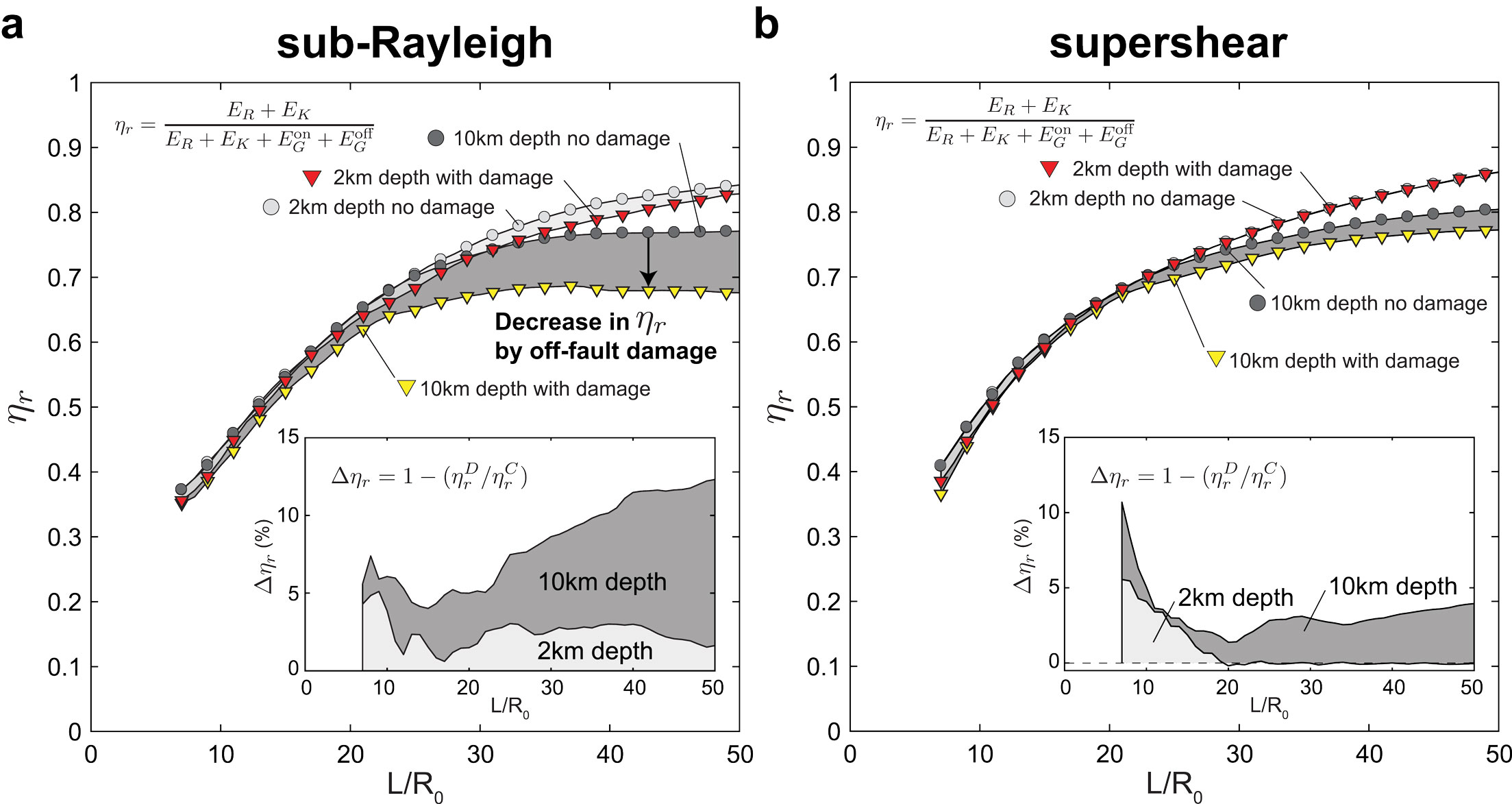}
\caption{Seismic efficiency at 2km and 10km depths with (a) $S$ $=$ $1.0$ (b) $S$ $=$ $0.7$ as a function of rupture length. The circles indicate the cases without  off-fault damage, whereas the inverted triangles indicate the cases with off-fault damage. The inset shows the percentage of the decrease in seismic efficiency due to the coseismic off-fault damage. Note that the rupture transitions to supershear around $L/R_0$ $=$ $40$ for the case without off-fault damage, while it remains sub-Rayleigh for the cases with off-fault damage in (a). The rupture transitions to supershear for both cases with and without off-fault damage in (b).}
\label{fig:seismicefficiency}
\end{figure} 

\begin{figure}
\center
\noindent\includegraphics[width=1.0\textwidth]{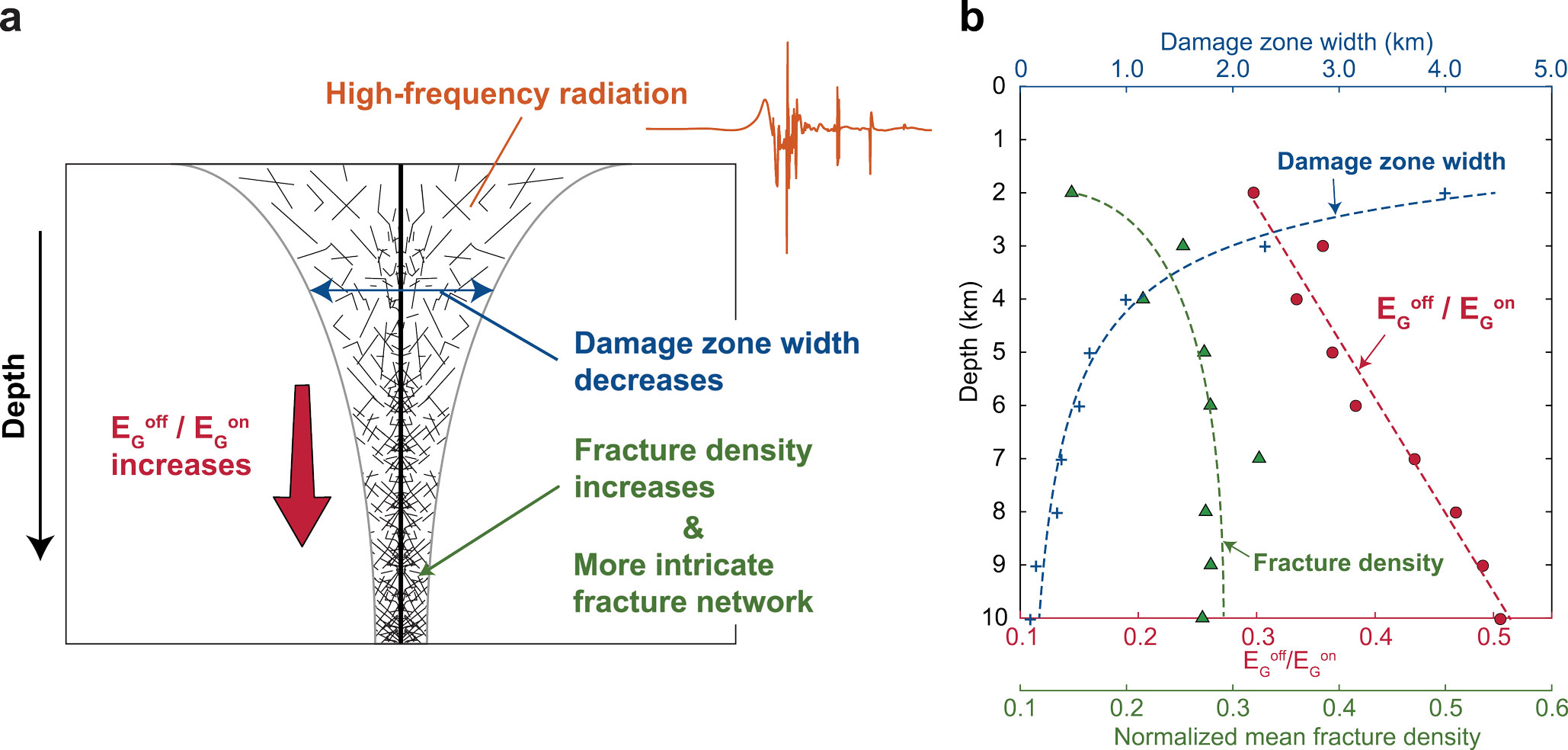}
\caption{Evolution of \remove{the} damage zone width, fracture density and contribution of off-fault damage to the overall energy budget. (a) Schematic of the off-fault damage with depth. (b) Damage zone width, fracture density and the ratio of dissipated fracture energy in the off-fault medium to the energy dissipated on the main fault with depth. The markers indicate the values at examined depths. Solid lines indicate expected trends of the discrete data.}
\label{fig:conclusivefigure}
\end{figure}

\clearpage

\renewcommand\thefigure{A.\arabic{figure}}
\setcounter{figure}{0}    

\begin{figure}
\center
\noindent\includegraphics[width=0.9\textwidth]{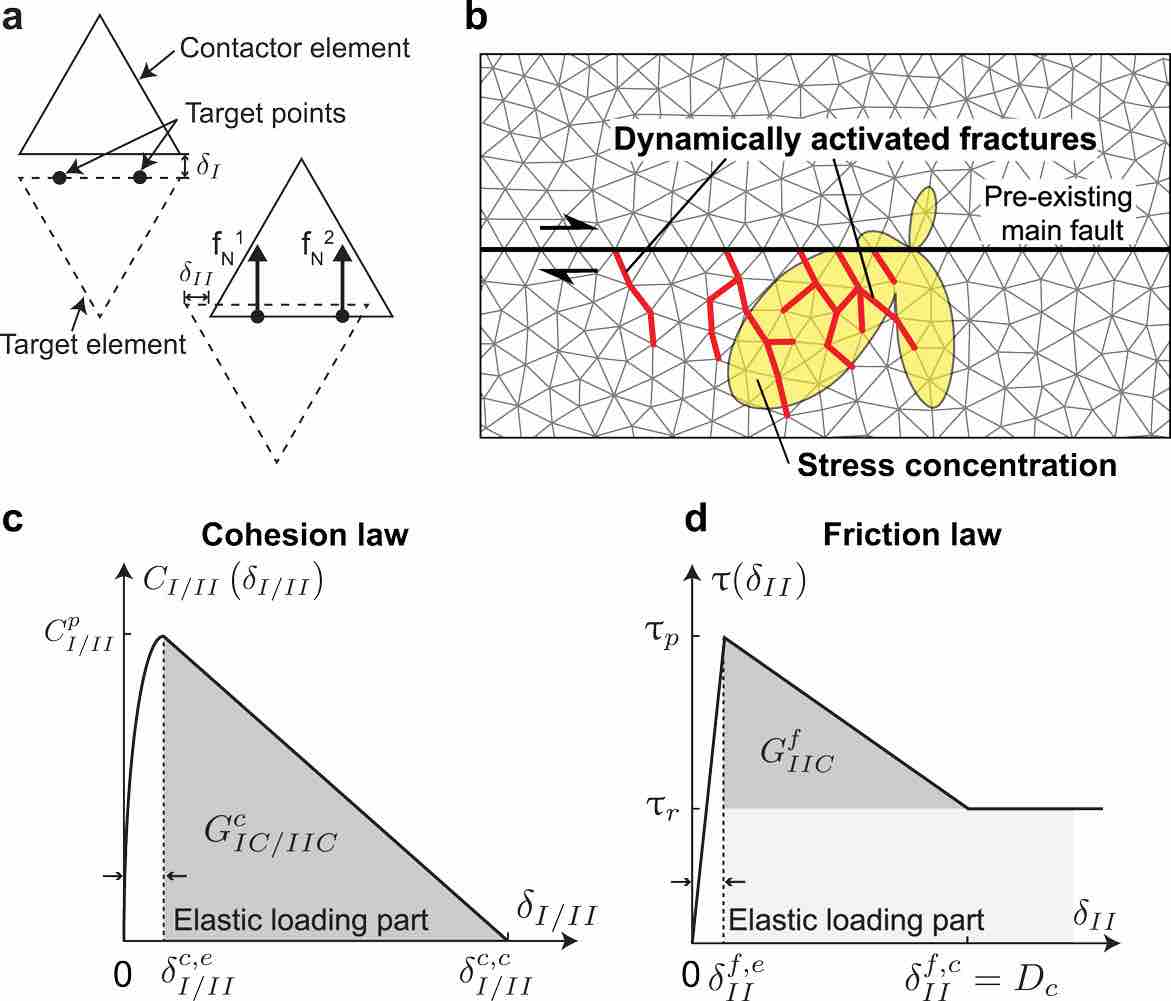}
\caption{Numerical framework of FDEM for dynamic earthquake rupture modeling. (a) Schematic of contactor and target. Opening displacement $\delta_I$, shear displacement $\delta_{II}$, and contact forces $f_N$ are indicated. The number of target points \change{drawn in}{indicated by} black dots is properly chosen for required numerical accuracy. (b) Model description showing the mesh discretization and the secondary fractures. Computational domain is discretized using an unstructured mesh. Every interface between elements is regarded as a potential failure plane, where cohesion and friction are operating as a function of $\delta_{I/II}$. When the tensile or shear cohesion starts weakening, we plot the interface as secondarily activated fractures as shown in red lines. (c) Linear displacement softening law for cohesion. The area highlighted in gray under the softening part of the curve indicates the fracture energy associated with cohesion in tension $G_{IC}^c$ and in shear $G_{IIC}^c$, respectively. (d) Linear slip-weakening law for friction. The energy dissipated by frictional process is divided into the fracture energy associated with friction $G_{IIC}^f$, while the rest is considered as heat energy.}
\label{fig:app1}
\end{figure}

\renewcommand\thetable{A.\arabic{table}}
\setcounter{table}{0}  

\begin{table}
\center
\small
\caption[Parameters used for Case Study with Depth.]{Parameters used for Case Study with Depth.}
\begin{tabular}{l l l}
\textbf{Variables} & &\textbf{Values} \\
\hline
$E$  $^a$ &Young's modulus & 75 GPa \\
$\mu$$^a$& Shear modulus & 30 GPa \\
$\nu$ $^a$& Poisson's ratio& 0.25 \\
$\rho$ $^a$& Density & 2700 kg m$^{-3}$ \\
$\rho_w$& Density of water & 1000 kg m$^{-3}$\\
$\psi$ & Orientation of $\sigma_1$ & 60 \\
$S$ &Seismic ratio& 0.7, 1.0 \\
$d_{MC}$ &Closeness to failure &0.4 \\
$f_s$  &Static friction coefficient & 0.6\\
$f_d$  &Dynamic friction coefficient & 0.2\\
$D_c$  & Critical slip distance & Estimated from eq. (\ref{eq:Dcwithdepth})\\
 $G_{IIC}^{f*}$ $^b$ & Fracture energy on the main fault& 3  MJ m$^{-2}$  \\
 $G_{IIC}^{f*, \text{off-fault}}$ $^b$ &Fracture energy in the off-fault medium& 0.01 MJ m$^{-2}$  \\
$C^p_{I}$ $^c$& Peak cohesion for tensile fractures & 8 MPa  \\
$C^p_{II}$ & Peak cohesion for shear fractures & Estimated from eq. (\ref{eq:CFII})\\
\hline \\
\label{tab:parameters}
\end{tabular}

\begin{flushleft}
\textbf{Note.} $^a$ Assuming representative values of granite \citep{nur1969}.  $^b$ \citet{viesca2015, passelegue2016b}. $^c$\citet{cho2003}.
\end{flushleft} 
\label{tab:parametersforcasestudy}
\end{table} 

\renewcommand\thefigure{B.\arabic{figure}}
\setcounter{figure}{0}    

\begin{figure}
\center
\noindent\includegraphics[width=0.9\textwidth]{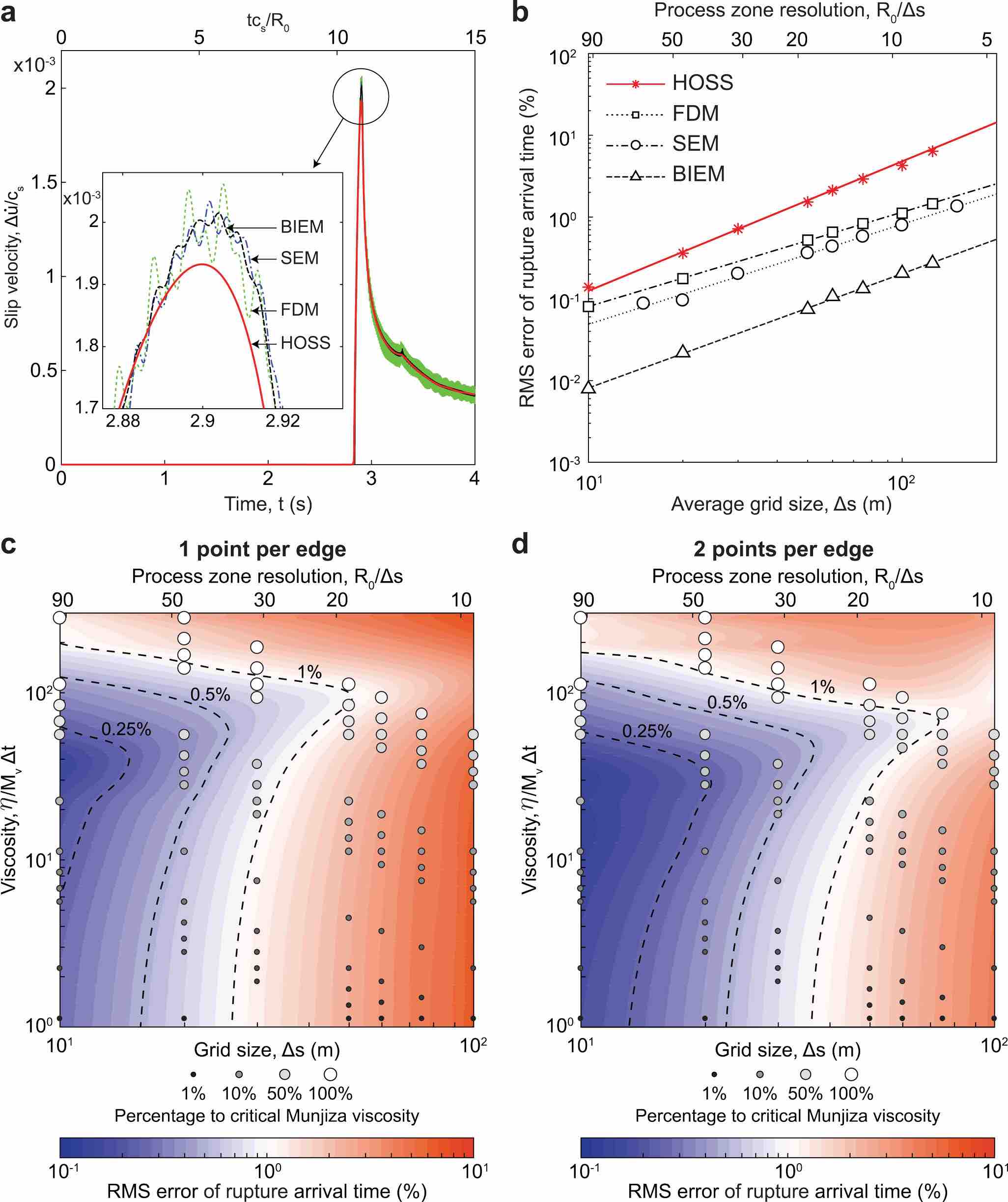}
\caption{Summary of the closs-validation of HOSS. (a) Slip velocity history at $x$ $=$ $9.0$ km ($x/R_0$ $=$ $9.7$). (b) Grid convergence as a function of grid size. The RMS error is calculated by the comparison of rupture arrival time to the benchmark result provided by \change{the highest-resolution solution of BIEM}{the solution of BIEM with highest-resolution}. The HOSS simulations are performed with two points per edge. (c, d) RMS error of the rupture arrival time with various combinations of viscosity and grid size with (c) one point and (d) two points per edge. The circles indicate the examined combinations, where the size of circles with monochromatic gradation represents the proportion of viscosity to the critical viscosity (the viscosity is higher with light color and \add{with} large circle). Color contour indicates the RMS error of rupture arrival time obtained by interpolating the examined combinations. }
\label{fig:cv1}
\end{figure}

\renewcommand\thefigure{C.\arabic{figure}}
\setcounter{figure}{0}    

\begin{figure}
\center
\noindent\includegraphics[width=0.9\textwidth]{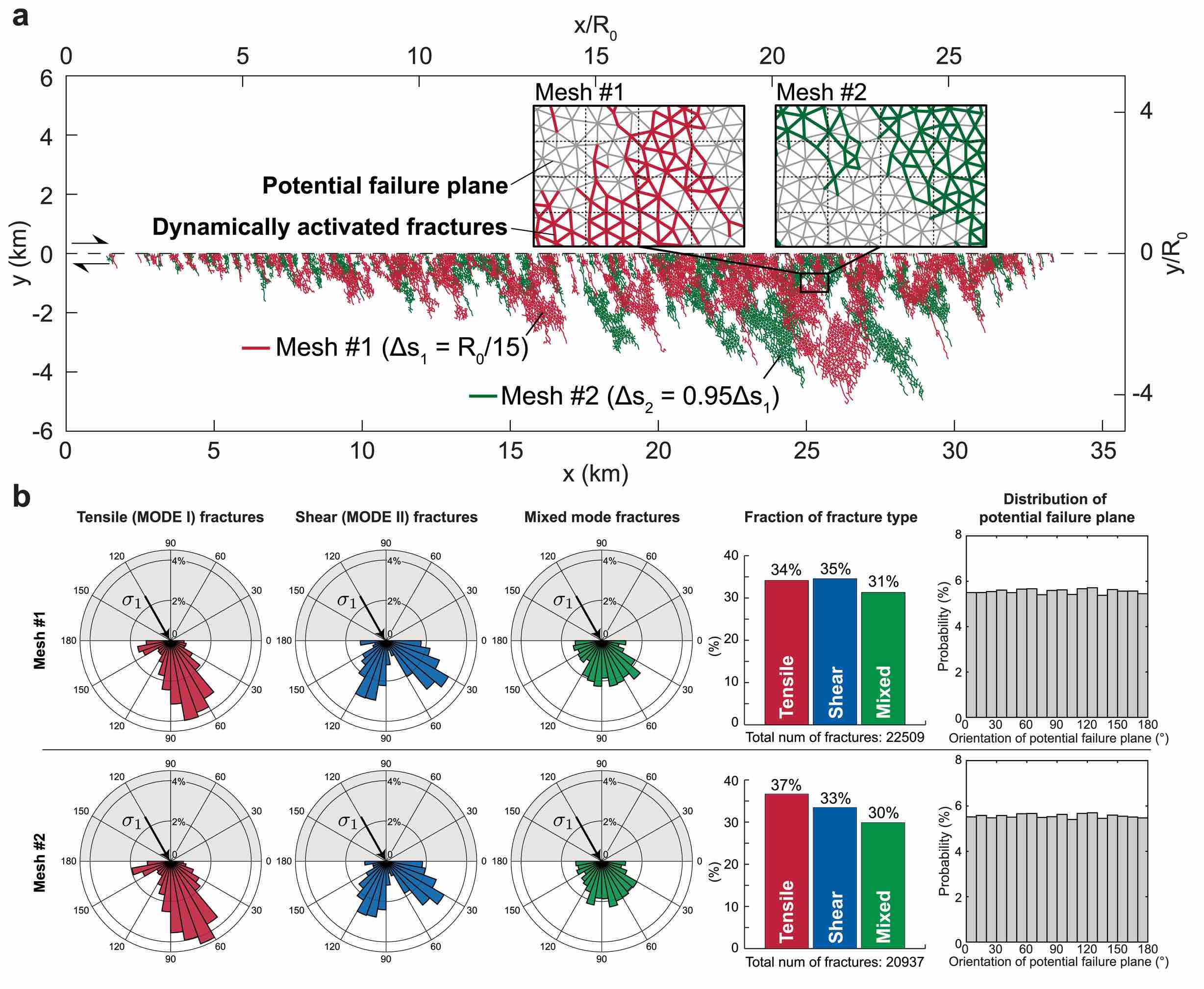}
\caption{Comparison of off-fault damage pattern. (a) Trace of off-fault fractures with different meshes. (b) Rose diagram of the orientation of off-fault fractures, fraction of damage type and distribution of potential failure planes.}
\label{fig:meshdepence_fracpattern}
\end{figure}

\begin{figure}
\center
\noindent\includegraphics[width=0.8\textwidth]{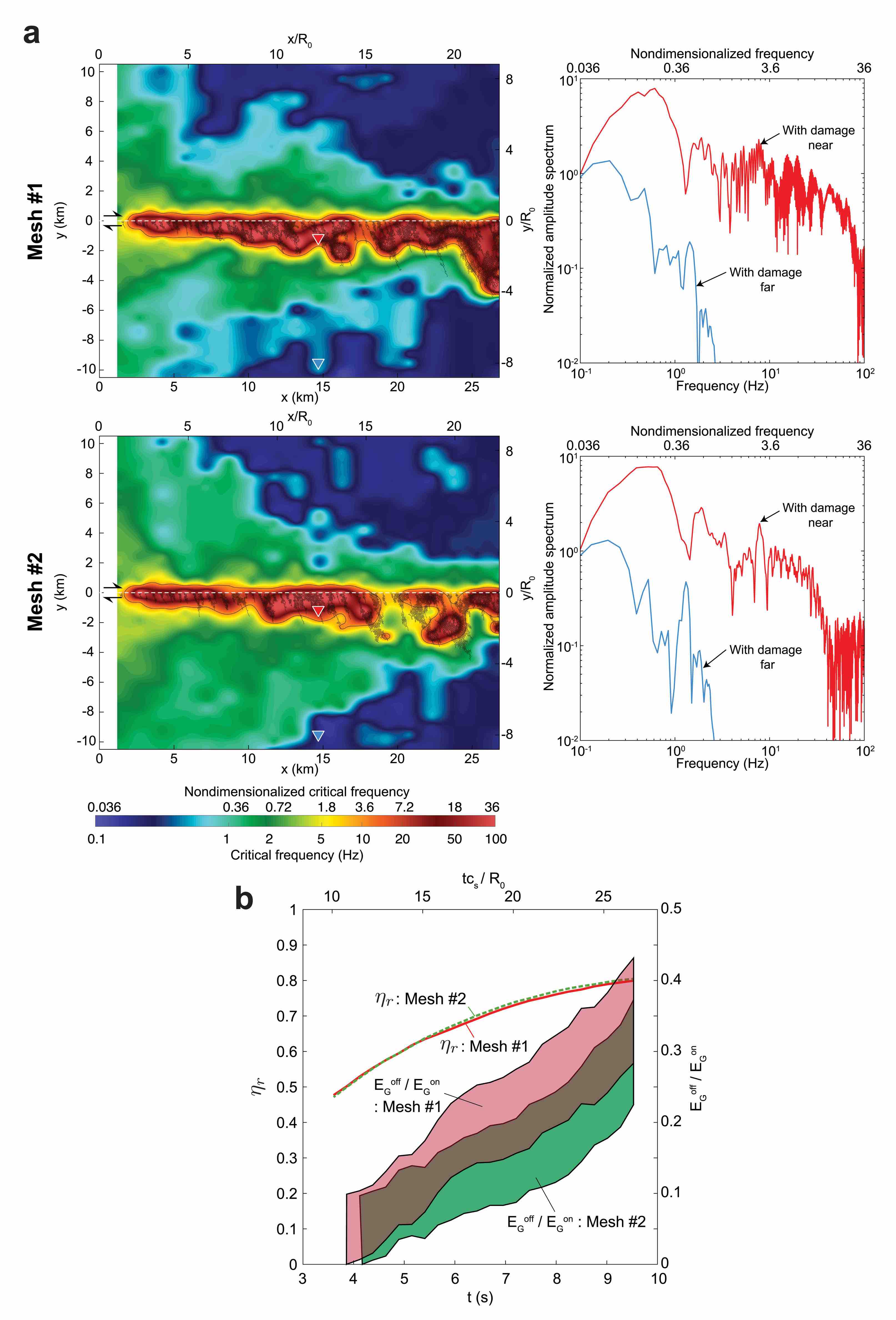}
\caption{Radiation and overall energy budget with different meshes. (a) Distribution of critical frequency in space and spectra at near and far from the main fault. (b) Seismic efficiency $\eta_r$ and the fraction of $E_G^{\text{off}}$ to $E_G^{\text{on}}$. The bands highlighted by color indicate the estimation of $E_G^{\text{off}}$/$E_G^{\text{on}}$ with the uncertainty of $\pm$ 15 \% in energy dissipation due to the numerical viscous damping.}
\label{fig:meshdepence_radiation}
\end{figure}

\end{document}